# Photooxidation of water with heptazine-based molecular photocatalysts: Insights from spectroscopy and computational chemistry

Wolfgang Domcke,[1] Andrzej L. Sobolewski,[2] and Cody W. Schlenker[3]

[1] Department of Chemistry, Technical University of Munich, D-85747 Garching, Germany

[2] Institute of Physics, Polish Academy of Sciences, PL-02-668 Warsaw, Poland

[3] Department of Chemistry, University of Washington, Seattle, Washington 98195, United States

**Abstract**

We present a conspectus of recent joint spectroscopic and computational studies which provided novel insight into the photochemistry of hydrogen-bonded complexes of the heptazine (Hz) chromophore with hydroxylic substrate molecules (water and phenol). It was found that a functionalized derivative of Hz, tri-anisole-heptazine (TAHz), can photooxidize water and phenol in a homogeneous photochemical reaction. This allows the exploration of the basic mechanisms of the proton-coupled electron-transfer (PCET) process involved in the water photooxidation reaction in well-defined complexes of chemically tunable molecular chromophores with chemically tunable substrate molecules. The unique properties of the excited electronic states of the Hz molecule and derivatives thereof are highlighted. The potential energy landscape relevant for the PCET reaction has been characterized by judicious computational studies. These data provided the basis for the demonstration of rational laser control of PCET reactions in TAHz-phenol complexes by pump-push-probe spectroscopy, which sheds light on the branching mechanisms occurring by the interaction of nonreactive locally excited states of the chromophore with reactive intermolecular charge-transfer states. Extrapolating from these results, we propose a general scenario which unravels the complex photoinduced water-splitting reaction into simple sequential light-driven one-electron redox reactions followed by simple dark radical-radical recombination reactions.

## 1. Introduction

During the past decade, polymeric carbon nitride materials, consisting of *s*-triazine or heptazine (tri-*s*-triazine) molecular building blocks, emerged as promising photocatalysts for the evolution of molecular hydrogen.[1-4] These materials are attractive because they consist of truly earth-abundant elements, are easily prepared from cheap precursors by standard procedures and are surprisingly stable under ultraviolet (UV) irradiation. The most widely employed material, generally referred to as graphitic carbon nitride (g-$C_3N_4$),[5] consists of imine-linked heptazine (Hz) units. Numerous modifications, exhibiting varying degrees of crystallinity or various kinds of morphology, chemical modification or doping with metal atoms, have been prepared and have been tested for hydrogen evolution efficiency under irradiation with UV light.[2-4] Unfortunately, the poorly constrained chemical compositions and atomic structures of these materials render the interpretation of the experimental data very difficult.

The vast literature on carbon nitrides mainly addresses materials properties, such as the band gap, the exciton dissociation efficiency, or the charge carrier mobility.[6-17] The photophysical properties of the Hz chromophore itself and the molecular aspects of the oxidation of water have received rather little attention so far. In this article, we describe an alternative perspective which focusses on the photochemical reactivity of molecular N-heterocycles with protic solvent molecules such as water or phenol. The research discussed herein has evolved from two originally independent developments, one in computational chemistry and the other in molecular spectroscopy and reaction kinetics. On the theoretical side, it was shown that photoexcited heterocycles, such as pyridine, triazine, or Hz, can oxidize water in hydrogen-bonded clusters via low-barrier excited-state proton-coupled electron-transfer (PCET) reactions and the detailed mechanisms of this photoreactivity were clarified.[18-20] On the experimental side, it was discovered that a functionalized derivative of Hz, tri-anisole-heptazine (TAHz), is chemically and photochemically remarkably stable and liberates OH radicals by photochemical hydrogen abstraction from liquid water.[21] TAHz also exhibits photocatalytic activity for $H_2$ evolution (with Pt as co-catalyst and with a sacrificial electron donor) which matches that of g-$C_3N_4$ in aqueous suspensions. Moreover, the initial steps of the photoreaction could be kinetically resolved by the measurement of dynamic luminescence quenching and the PCET character of the photoreaction was confirmed by the detection of the kinetic isotope effect in $H_2O$ *vs.* $D_2O$.[21] Taken together, these theoretical and experimental findings suggest that the basic mechanisms of the water photooxidation reaction can be explored by the investigation of precisely defined and chemically tunable molecular chromophores in neat solvents. The harrowing ambiguities of photochemistry in poorly defined complex polymeric materials can thereby be avoided.

While the solubility of TAHz in liquid water is low, the molecular photophysics of TAHz can more conveniently be investigated in organic solvents, for example in toluene. Replacing water by phenol (substituted water), which also is readily soluble in organic media, can serve as a model molecular testbed in which photoinduced PCET reactions between hydrogen-accepting chromophores and hydroxylic reaction partners can be studied under precisely controlled conditions. While phenol exhibits a significantly lower oxidation potential than water (and therefore is a sacrificial electron donor), its oxidation potential can be tuned over a wide range by functional groups.[22-26] The effect of the oxidation potential (as well as other properties) of the electron donor on the hydrogen evolution reaction can therefore systematically be investigated. This strategy apparently was not considered so far in the context of photocatalytic hydrogen evolution with solid-state organic or inorganic photocatalytic materials.

Since toluene is an inert solvent, its effect on the photochemistry of the chromophore-phenol complexes should be minor. The interpretation of the spectroscopic and kinetic studies can therefore be supported by *ab initio* electronic-structure calculations for the isolated chromophore-phenol complexes. The ab initio electronic-structure calculations referred to in this article were performed with the ADC(2) method, which is a computationally efficient wave-function-based single-reference propagator method.[27-30] It yields electronic excitation energies with an accuracy of a few tenths of an electron volt and provides, in particular, a balanced description of locally excited electronic states and electronic charge-transfer (CT) states in molecular complexes, which is an essential feature for accurately describing PCET reactions. In previous work, the ADC(2) method was benchmarked against multi-configuration multi-reference methods (CASSCF/CASPT2) for complexes of small heterocycles (pyridine, triazine) with water molecules.[18, 19] For a more detailed description of the computational methods, the reader is referred to these earlier publications.

The structure of this perspective article is as follows. First, the unusual and rather unique spectroscopic properties of the isolated Hz molecule and of derivatives of Hz are analysed with ab initio electronic-structure calculations. The photochemical reactivity is explored with computational and experimental methods for hydrogen-bonded complexes of Hz-based chromophores with protic substrate molecules. The complexes form spontaneously by self-assembly in solution. In addition to the locally excited electronic states of the chromophore, there exists an intermolecular CT state which involves electron transfer from the substrate molecule to the Hz molecule. The calculations reveal that the energy of the CT state can be tuned over a wide range by functional groups on either the Hz molecule or on the substrate. The energy of the CT state, in turn, controls the barrier of the PCET reaction. It is shown by the interplay of theory and experiment that the PCET reaction can be chemically tuned from a barrier-dominated reaction to a barrierless reaction on the potential energy (PE) surface of the lowest excited singlet state. A specific feature of the Hz chromophore, the exceptionally long lifetime of the lowest excited singlet state, allows the demonstration of active laser control of the PCET reaction by pump-push-probe spectroscopy, which yields deep insight into the interplay of photophysical and photochemical reaction mechanisms. In addition to radical pair formation by PCET, radical recombination reactions have been explored with computational methods. It is shown that under certain conditions Hz-based photocatalysts have the capability of spontaneous self-repair after undesired radical recombination reactions. Finally, the closure of the photocatalytic cycle is discussed. The reduced Hz radicals (HzH) generated by the sequential homolytic decomposition of two $H_2O$ molecules can recombine in thermal radical-radical recombination reactions, which yields molecular hydrogen and regenerates the Hz photocatalyst. Alternatively, the Hz chromophore can be recovered by photodetachment of the excess hydrogen atom from the HzH radical, resulting in the release of free H-atom radicals (in nonpolar solvents) or solvated electrons (in aqueous solution).

## 2. Excited electronic states of Hz-based chromophores

### 2.1. Excited states of Hz

Hz exhibits $D_{3h}$ symmetry. The highest occupied molecular orbital (HOMO) and the lowest unoccupied molecular orbital (LUMO) are nondegenerate $\pi$ orbitals which are displayed in Fig. 1. The lowest excited singlet state of Hz, $S_1(A_2')$ with a calculated vertical excitation energy of 2.60 eV at the ADC(2)/cc-pVDZ level, is a nearly pure HOMO-LUMO excitation. The dipole transition moment from the electronic ground state is zero by symmetry. The next higher $\pi\pi^*$ excited singlet state, $S_4(E')$ with an excitation energy of 4.43 eV at the ADC(2)/cc-pVDZ level, results from the excitation of an electron from the HOMO to the degenerate $\pi$ orbital above the LUMO. This transition carries significant oscillator strength ($f$ = 0.54). The $S_4$ state is responsible for the onset of strong absorption in Hz-based condensed materials near 400 nm. Between the $S_1(\pi\pi^*)$ and $S_4(\pi\pi^*)$ states there exist two $n\pi^*$ excited singlet states, $S_2(A_1'')$ and $S_3(E'')$, see Table 1. The $^1n\pi^*$ states are electronically dark, but may borrow intensity from the nearby bright $^1\pi\pi^*$ state by vibronic coupling. Experimental data on the electronic excitation spectrum of the isolated Hz molecule are not available because Hz hydrolyses rapidly under ambient conditions.[31, 32]

**Table 1.** Vertical excitation energies (in eV) of the lowest four singlet and triplet excited states of Hz, calculated with the ADC(2) method with the cc-pVDZ basis set. Oscillator strength for singlet transitions is given in parentheses.

| $S_1\ A_2'(\pi\pi^*)$ | 2.57 (0.0) | $T_1\ A_2'(\pi\pi^*)$ | 2.85 |
|---|---|---|---|
| $S_2\ A_1''(n\pi^*)$ | 3.76 (0.0) | $T_2\ E'(\pi\pi^*)$ | 3.67 |
| $S_3\ E''(n\pi^*)$ | 3.85 (0.0) | $T_3\ A_1''(n\pi^*)$ | 3.76 |
| $S_4\ E'(\pi\pi^*)$ | 4.43 (0.538) | $T_4\ E''(n\pi^*)$ | 3.82 |

The triplet manifold exhibits the same pattern of nondegenerate and degenerate $\pi\pi^*$ and $n\pi^*$ excited states, see Table 1. Inspection of Table 1 reveals, however, a very unusual feature of the triplet spectrum of Hz. While the $^3E'$ state ($T_2$) is located 0.76 eV below the corresponding $^1E'$ state, in accordance with common expectation, the $^3A_2'$ state ($T_1$) is located 0.28 eV *above* the $^1A_2'$ state ($S_1$). The $S_1/T_1$ energy gap, defined as $\Delta_{ST} = E_{S1} - E_{T1}$, thus is inverted (negative). This feature was so far unknown for any aromatic molecule. Very recently, the $S_1/T_1$ energy inversion was computationally established also for the related heterocycle cycl[3.3.3]azine.[33]

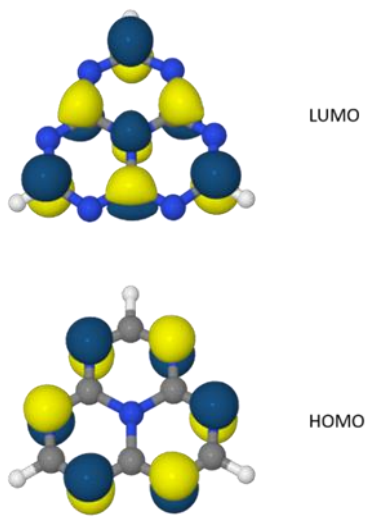

*Fig. 1. Highest occupied (HOMO) and lowest unoccupied (LUMO) Hartree-Fock molecular orbital of Hz.*

The origin of the $S_1/T_1$ inversion in Hz can be qualitatively understood by the inspection of Fig. 1. While the HOMO is exclusively localized on the peripheral nitrogen atoms, the LUMO is localized on the carbon atoms and the central nitrogen atom. This peculiar pattern results in zero spatial overlap of HOMO and LUMO. As a consequence, the exchange integral $K$, which primarily determines the singlet-triplet energy gap, is strongly suppressed. For Hz, $K = 0.12$ eV, which is very small compared with the Coulomb integral, $J = 5.62$ eV. Coupling with double excitations, which can be interpreted as spin polarization in the unrestricted orbital picture,[34] stabilizes the singlet state relative to the triplet state, resulting in the inverted $S_1/T_1$ energy gap.[35]

Exploratory computational results indicate that $S_1/T_1$ inversion is a very robust property of Hz chromophores. Neither substitution of Hz nor oligomerization nor stacking has any significant effect on the $S_1/T_1$ energy gap.[35] These results provide strong evidence that $S_1/T_1$ energy inversion exists also in crystalline and polymeric Hz-based materials.

The theoretical prediction of $S_1/T_1$ inversion was carefully scrutinized by spectroscopic investigations for the TAHz chromophore.[35] The photoluminescence decay of TAHz in toluene was measured with high precision on the microsecond timescale. The effects of molecular oxygen (a triplet quencher) and of ethyl iodide (a heavy atom carrier) was investigated. No evidence of triplet quenching was found, nor could phosphorescence be detected. Microsecond transient absorption experiments did not find evidence of the computationally predicted triplet-triplet absorption, whereas the predicted singlet-singlet absorption was readily detected.[35] These data convincingly confirm the inversion of the $S_1$ and $T_1$ states in TAHz. The photophysics of several analogues of Hz, such as cycl[3.3.3]azine and

triazacycl[3.3.3]azine, was explored earlier by Wirz and coworkers.[36, 37] They found evidence for unusual luminescence properties and singlet-triplet near-degeneracy, but could not unequivocally establish singlet-triplet inversion.

The singlet-triplet inversion has important implications for the photochemistry of the molecular Hz chromophore and of Hz-based materials. The absence of a triplet state below the lowest singlet state obviously eliminates the decay channel of intersystem crossing (ISC) from the lowest singlet state. The absence of quenching by ISC together with the symmetry-forbidden radiative decay endows the $S_1$ state of Hz with an exceptionally long lifetime. For TAHz in toluene, the $S_1$ lifetime has been established as 287 ns with a fluorescence quantum yield of 0.69.[21] As will be discussed below, the $S_1$ state plays the role of a reservoir state in the photophysics of Hz chromophores in which a significant part of the energy of the absorbed photon is stored in the excited state on time scales which are relevant for molecular diffusion and intermolecular reactions.

It is well established that aza-arenes generally possess low-lying long-lived triplet states which, upon irradiation, are populated with high quantum yields by ISC. These long-lived triplet states efficiently generate reactive singlet oxygen ($^1O_2$) from atmospheric triplet oxygen ($^3O_2$). The generation of reactive $^1O_2$ is a long-standing problem in artificial photosynthesis with organometallic photocatalysts.[38, 39] The $S_1/T_1$ inversion in Hz chromophores completely eliminates this problem for Hz-based materials. Indeed, unexpected ultralow yields of singlet oxygen have been reported for irradiated g-$C_3N_4$[40-42] and may be one of the reasons for the exceptional photostability of these materials.

**2.2. Excited states of Hz derivatives: tuning the absorption spectrum**

TAHz is a multi-chromophoric system. The excitonic interactions between the excited states of the Hz core and the excited states of the anisole pendant groups have a substantial impact on the absorption spectrum and the absorption cross section. It is instructive to take a more systematic look at this phenomenon by considering substitutions of Hz with a selection of aromatic pendants. The vertical excitation energies and oscillator strengths of the five lowest singlet excited states of the Hz derivatives tri-phenyl-heptazine (TPhHz), tri-pyrazinyl-heptazine (TPyHz) and tri-anisole-heptazine (TAHz) are listed in Table 2. These data may be compared with the data for Hz in Table 1. For the substituted heptazines, a pair of bright $^1\pi\pi^*$ states is found near 4.0 eV. The energy of the lowest bright $^1\pi\pi^*$ state is lowered by about 0.38 eV in TPhHz, 0.42 eV in TPyHz, and by 0.80 eV in TAHz relative to Hz. The cumulative oscillator strength of the pair of lowest absorbing states increases from 2.389 for TPhHz and 2.417 TPyHz to 2.657 for TAHz. It is obvious that TAHz is a significantly improved light-harvesting chromophore compared with Hz. The larger effect of the anisole substituent compared with benzene and pyrazine arises from the electron donating character of this moiety.

Remarkably, the energy of the low-lying $S_1(\pi\pi^*)$ state of Hz is marginally affected by the substituents, see Table 2.

**Table 2**. Vertical excitation energies (in eV) and oscillator strengths (in parentheses) of the five lowest singlet excited states of tri-phenyl-heptazine (TPhHz), tri-pyrazinyl-heptazine (TPyHz) and tri-anisole-heptazine (TAHz).

| State | TPhHz | TPyHz | TAHz |
|---|---|---|---|
| $^1A(\pi\pi^*)$ | 2.55 (0.0) | 2.39 (0.0) | 2.62 (0.0) |
| $^1A(n\pi^*)$ | 3.58 (0.0) | 3.27 (0.0) | 3.66 (0.0) |
| $^1E(n\pi^*)$ | 3.70 (0.0) | 3.36 (0.0) | 3.79 (0.0) |
| $^1E(\pi\pi^*)$ | 4.05 (1.910) | 4.01 (0.952) | 3.62 (2.378) |
| $^1E(\pi\pi^*)$ | 4.22 (0.479) | 4.39 (1.465) | 4.17 (0.279) |

As an example of a series of extended-chain aromatic pendant groups, Fig. 2 shows the calculated absorption spectra of Hz decorated with a chain of n pyrazine molecules, n = 1 … 5. For reasons of computational cost, only a single chain of pyrazine molecules is considered. It can be seen that the peak of the absorption spectrum is shifted to lower energies from ≈ 4.4 eV (280 nm) for n = 1 to ≈ 3.6 eV (340 nm) for n = 5. The oscillator strength increases by a factor of three from n = 1 to n = 5. The absorption spectrum and the absorption cross section of the Hz chromophore can thus be substantially tuned by extended chains of aromatic groups.

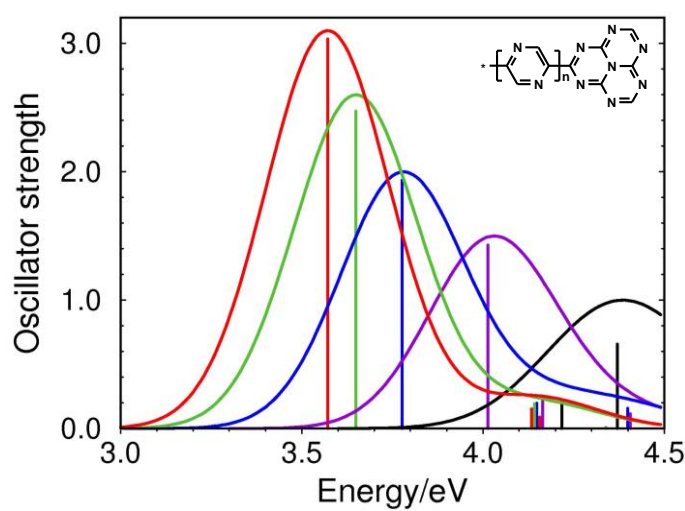

*Fig. 2. Calculated absorption spectra of Hz decorated with a chain of n pyrazine molecules: n = 1 - black, n = 2 - violet, n = 3 - blue, n = 4 - green, n = 5 - red. The computed stick spectra were convoluted with a Gaussian function of FWHM = 0.4 eV.*

A more complete picture of the vertical excitation spectrum of Hz functionalized with chains of pyrazine molecules is given in Fig. 3. The energies of the bright ($^1E$) and dark ($^1A$) singlet states and the corresponding triplet states ($^3E$, $^3A$) are displayed as functions of the chain length. Concordant with the lowering of the energy of the bright $^1\pi\pi^*$ state ($^1E$) of Hz by aromatic functionalization, the corresponding $^3\pi\pi^*$ state ($^3E$) comes down in energy. At a certain chain length, the energy of the $^3E$ state may drop below the energy of the ¹A state. When this happens, the singlet-triplet energy inversion of Hz is removed and the chromophore will possess a long-lived (phosphorescent) triplet state. It also has to be kept in mind that the $^3E$ state is spatially degenerate in $D_{3h}$ symmetry and therefore exhibits, in contrast to the nondegenerate $S_1$ and $T_1$ states, a Jahn-Teller effect which may substantially lower the minimum energy of this state by the Jahn-Teller stabilization energy.

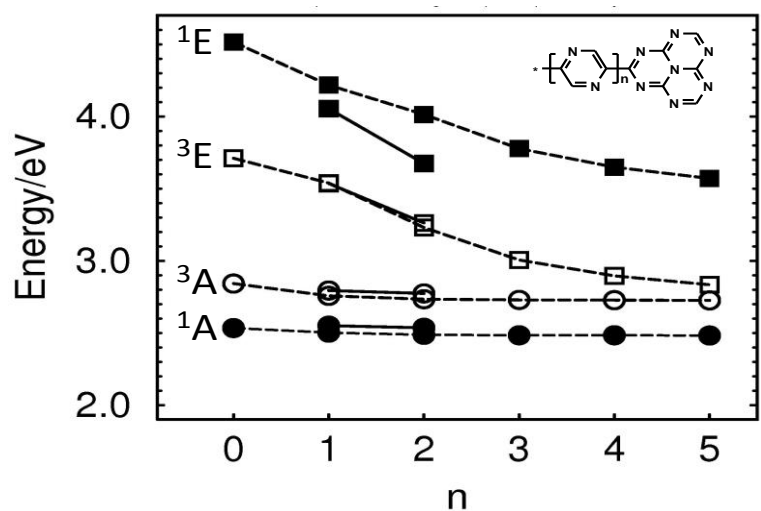

*Fig. 3. Vertical excitation energies of the bright ($^1E$) and dark ($^1A$) $^1\pi\pi^*$ states and the corresponding $^3\pi\pi^*$ states of Hz functionalized with chains on n pyrazine molecules, n = 1 ... 5. The symbols connected by dashes lines represent substitution by a single chain of pyrazine molecules. The symbols connected by full lines (n = 1 and n = 2) represent triply substituted Hz. It can be seen that the effects of single and triple substitution are similar, except for the $^1E$ state, which is stabilized more strongly by triple substitution.*

By the energy gap law, the lowering of the energy of the absorbing $^1\pi\pi^*$ state in aromatically substituted heptazines enhances the radiationless deactivation rate of this state to the long-lived $S_1(\pi\pi^*)$ state. The ligands thus affect the competition between radiationless decay and reactive PCET processes in complexes of these chromophores with protic substrate molecules.

## 3. Photoinduced PCET in hydrogen-bonded complexes of Hz with protic substrates

### 3.1. The Hz-H₂O complex

Hz has six peripheral nitrogen atoms which can serve as acceptor atoms for hydrogen bonding with a water molecule. The structure of the Hz···H₂O complex with in-plane orientation of the water molecule is shown in Fig. 4. The length of the hydrogen bond is 2.046 Å at the MP2/cc-pVDZ computational level, which represents a relatively strong hydrogen bond. A closer analysis reveals that the nonbonding orbital of the accepting N-atom is partially delocalized over the $p_{x,y}$ orbitals of the oxygen atom of the water molecule, indicating a partially covalent "non-innocent" hydrogen bond.[20]

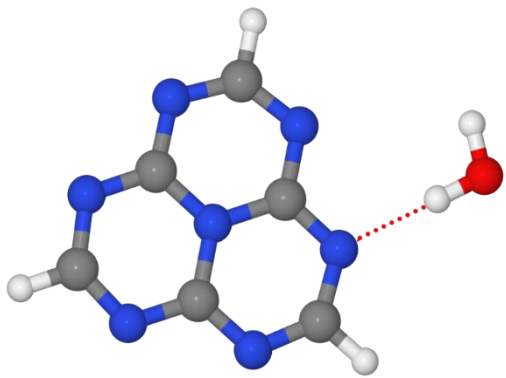

*Fig. 4. Ground-state equilibrium geometry of the heptazine-$H_2O$ complex. The hydrogen bond is indicated by the dotted line.*

The HOMO and LUMO of the complex are the π and π* orbitals of Hz. They are practically indistinguishable from those shown in Fig. 1. The electronic excitation spectrum of the complex is essentially identical with that of isolated Hz. The degeneracy of the $^1$E'(ππ*) and $^1$E"(nπ*) states of isolated Hz is lifted, but the energy splittings are less than 0.1 eV.[20]

We define the bond length $R_{OH}$ of the OH group of the water molecule involved in the hydrogen bonding (see Fig. 4) as the reaction coordinate for the H-atom transfer reaction from the water molecule to Hz. For small displacements $R_{OH}$, the potential-energy (PE) functions of the $^1$ππ* and $^1$nπ* excited states of the complex are essentially parallel to the PE function of the electronic ground state, as can be seen on the left hand side of Fig. 5, revealing that the OH bond of the water molecule is little affected by the local electronic excitation of Hz.

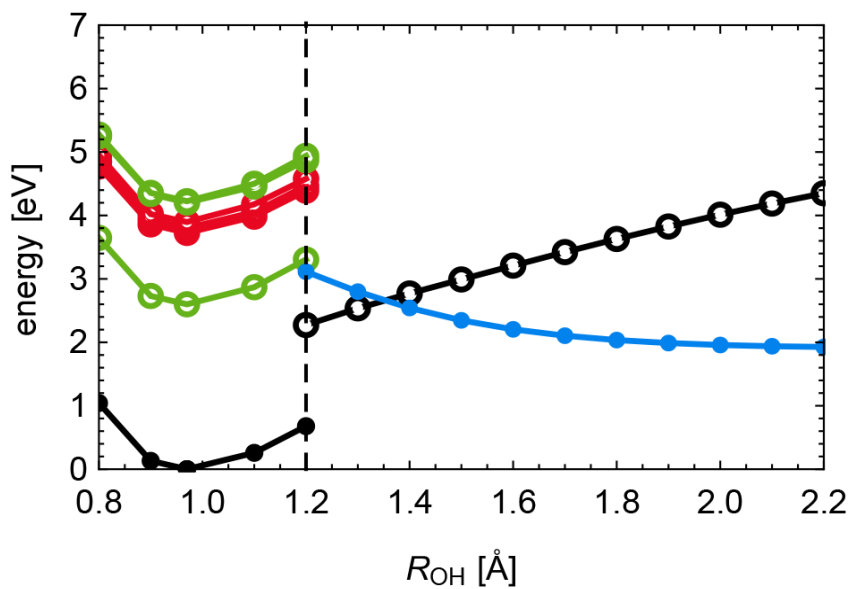

*Fig. 5. Energy profiles of the electronic ground state and the lowest excited states of the heptazine-$H_2O$ complex along minimum-energy paths for H-atom transfer from water to heptazine, calculated with the ADC(2) method. Full circles indicate that the reaction path has been optimized in this state. Open circles represent the energies of electronic states which have been calculated for geometries optimized in a different electronic state. The dashed vertical line separates the reaction path optimized in the $S_0$ state (left) from the reaction path optimized in the lowest charge-transfer state (right). Black: $S_0$ state; green:*

*locally excited $^1\pi\pi^*$ states, red: locally excited $^1n\pi^*$ states; blue: charge-transfer state. [Adapted with permission from J. Ehrmaier et al., J. Phys. Chem. A **121**, 4754 (2017). Copyright American Chemical Society 2017].*

This is, however, no longer true when large changes of $R_{OH}$ or of the distance $R_{ON}$ of the donor oxygen atom and the acceptor nitrogen atom are considered. For extended $R_{OH}$ or reduced $R_{ON}$, an additional electronic state becomes relevant which arises from the excitation of an electron from the $p_z$ orbital on the O-atom of water to the $\pi^*$ orbital of Hz. This excited state is located at high energy (5.40 eV) at the ground-state equilibrium geometry of the Hz⋯H$_2$O complex. When the energy of this state is minimized by geometry optimization, this state not only becomes the lowest excited state of the complex, but its energy even drops below the energy of the closed-shell ground state. This phenomenon is illustrated on the right hand side of Fig. 5. The crossing of the energy of the charge-transfer (CT) state (blue dots) with the energy of the closed-shell state (black circles) represents a conical intersection[43] of two electronic states of the same symmetry ($^1$A'). The minimum of the PE surface of the CT state represents a new chemical species which is a hydrogen-bonded HzH⋯OH biradical, see Scheme 1(a). It can be inferred from Fig. 5 that about 2.0 eV of the energy of the absorbed photon (≈ 4.2 eV) is stored as chemical energy in the biradical.

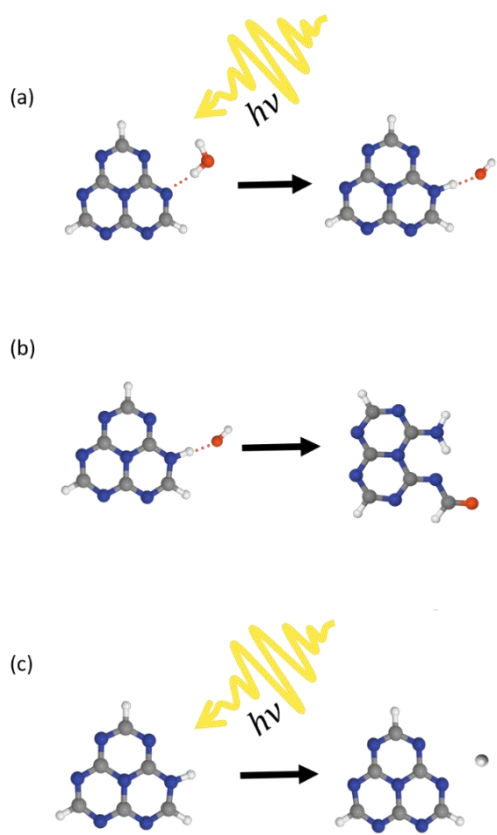

*Scheme 1. (a) A photon triggers the H-atom transfer reaction from water to the Hz chromophore, forming the HzH⋯OH radical pair. (b) The HzH and OH radicals can recombine in a dark reaction to form photohydrates of Hz. (c) Photodetachment of the excess hydrogen atom from the HzH radical.*

The PE functions shown on the left hand side and right hand side of Fig. 5 belong to differently optimized nuclear geometries and therefore do not reveal the barrier between the left and the right minima. The barrier becomes apparent when a two-dimensional relaxed PE surface of the S$_1$ state is constructed as a function of the H-atom transfer coordinate $R_{OH}$ and the donor-acceptor distance $R_{ON}$. This surface, which is displayed in Fig. 6, exhibits a topography which is characteristic for H-atom transfer reactions.[44-49] The well on the left hand side, corresponding to the locally excited $^1\pi\pi^*$ state of Hz, and the valley on the right hand side, corresponding to the HzH···OH biradical, are separated by a barrier of 0.75 eV relative to the $S_1(\pi\pi^*)$ minimum. While the $S_1$ state is extremely long-lived in the isolated Hz molecule, as discussed above, its population can be quenched in the Hz···H$_2$O complex by H-atom transfer, either by over-the-barrier dynamics if sufficient vibrational excess energy is available, or by H-atom tunnelling through the barrier.

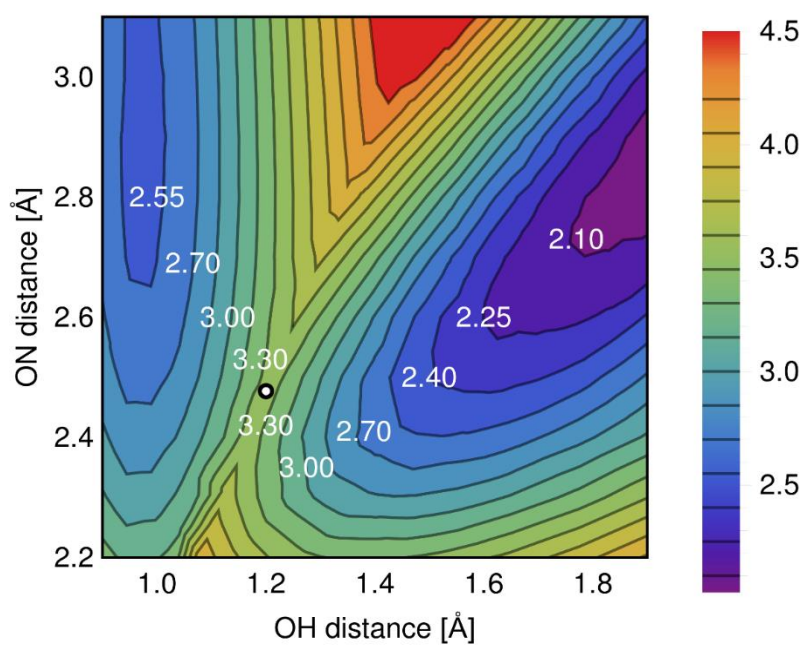

*Fig. 6.* PE surface of the lowest excited singlet state in the vicinity of the barrier for H-atom transfer from water to heptazine in the heptazine-H$_2$O complex, calculated with the ADC(2) method. The nuclear coordinates are the OH bond length of water and the distance of the oxygen atom of water from the peripheral nitrogen atom of heptazine. The PE surface is relaxed with respect to all other internal coordinates of the complex (except the CH bond lengths of heptazine). The numbers give the potential energy relative to the ground-state minimum in electron volts. The circle indicates the location of the saddle point. [Adapted with permission from J. Ehrmaier et al., J. Phys. Chem. A **121**, 4754 (2017). Copyright American Chemical Society 2017].

The high-energy region (orange) near the upper rim of Fig. 6 represents the lower cone of a conical intersection between the $S_1(\pi\pi^*)$ state and the CT state. The minimum-energy path of the PCET reaction on the $S_1$ PE surface bypasses this conical intersection and the character of the electronic wave function changes gradually (adiabatically) from $^1\pi\pi^*$ to CT along the reaction path. The PCET in the Hz···H$_2$O complex therefore is an adiabatic photoreaction. However, a seam of intersection of the PE surface of the biradical and the PE surface of the closed-shell state exists, as can be seen on the right hand side of Fig. 5. Along this seam of intersection, a nonadiabatic reaction (bifurcation of the nuclear wave packet) takes place.

Experimentally, the water photooxidation reaction was explored using suspensions of TAHz in water.[21] The extinction and stationary photoluminescence spectra of TAHz in water are shown in Fig. 7(a). The absorption of the bright $^1\pi\pi^*$ state exhibits an onset at 400 nm. Upon closer inspection, three weak peaks are visible near the red edge of the absorption spectrum, see Fig. 7(b). The two more intense peaks, which are absent for TAHz in toluene, arise from $^1n\pi^*$ states which are blue-shifted in water due to hydrogen bonding and acquire intensity due to their proximity to the bright $^1\pi\pi^*$ state. The photoluminescence in Fig. 7(a) originates from the nominally dark long-lived $S_1(\pi\pi^*)$ state. The lifetime of the $S_1$ state of TAHz is reduced from 287 ns in toluene to 16 ns in water.[21] In addition to the slow luminescence from the $S_1$ state, a faster blue-shifted luminescence could be detected which decays on a timescale of 190 ps with a kinetic isotope effect (KIE) of 2.9 in $H_2O$ vs. $D_2O$.[21] This "non-Kasha" luminescence originates from the bright $^1\pi\pi^*$ state of TAHz. The KIE confirms that the quenching occurs by proton transfer from water. The 190 ps decay time reveals that PCET in the higher $^1\pi\pi^*$ state is considerably faster than in the $S_1(\pi\pi^*)$ state, which can be explained by a lower PCET barrier in the higher $^1\pi\pi^*$ state.[20] In addition, it was shown that irradiation of TAHz in aqueous solution with 365 nm LED light generates free OH radicals which were detected by scavenging with terepthalic acid[21] as well as by spin-trapping EPR spectroscopy, see Fig. 8.

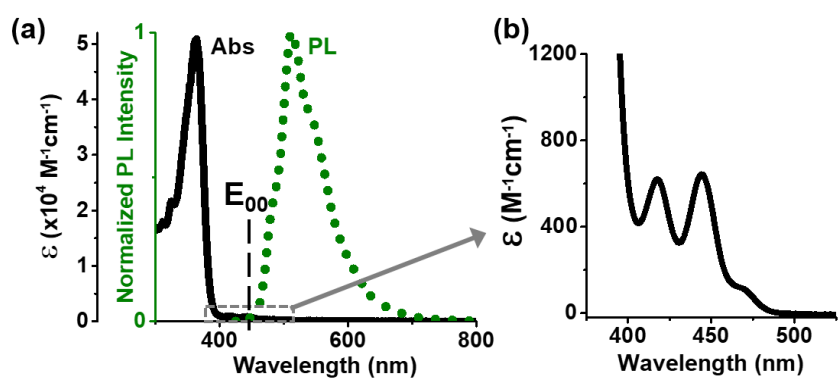

*Fig. 7. (a) TAHz optical extinction spectrum (solid, $\lambda_{max}$=365 nm) and photoluminescence spectrum (dotted, $\lambda_{max}$=505 nm) for a 33 μM toluene solution. (b) Extinction spectrum of weakly-allowed features below the absorption band of the bright $\pi\pi^*$ state. [Adapted with permission from E. J. Rabe et al., J. Phys. Chem. Lett. 9, 6257 (2018). Copyright American Chemical Society 2018].*

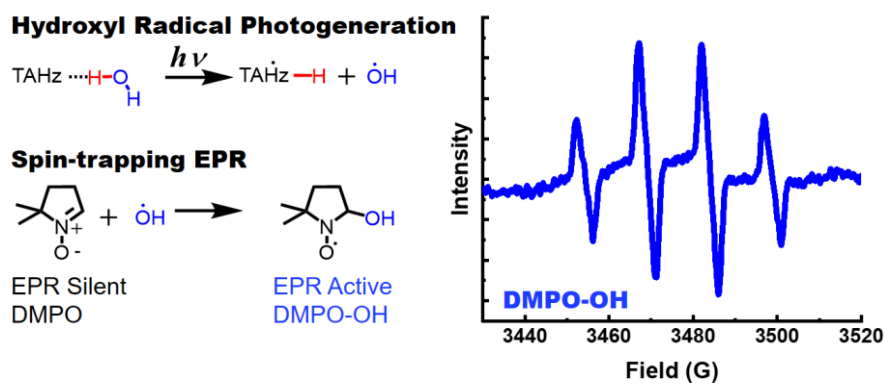

*Fig. 8. Photoinduced H-atom abstraction from water by TAHz under 365 nm illumination yields hydroxyl radicals. The hydroxyl radicals can be effectively scavenged using a spin-trapping reagent such as 5,5-dimethyl-1-pyrroline N-oxide (DMPO) to form a long-lived radical adduct (DMPO-OH), which can be detected by EPR, DMPO itself being EPR silent. The EPR spectrum on the right shows the diagnostic signature of the photogenerated DMPO-OH adduct that is formed as a result of TAHz illumination in water.*

These experimental data provide clear and reproducible evidence that the photoexcited TAHz molecule indeed can oxidize water, not just sacrificial electron donors with low oxidation potentials. Moreover, it has thereby been shown that a Hz-based chromophore in solution can oxidize water in a homogenous photochemical reaction, providing evidence that water oxidation photocatalysis is not *a priori* dependent on charge separation and charge transport processes in solid-state materials.

### 3.2. Complexes of TAHz with functionalized phenols

Unlike water, phenol is soluble in organic solvents. Given the good solubility of TAHz in toluene, hydrogen-bonded complexes of TAHz with phenol in toluene solution represent ideal model systems for the exploration of the photochemistry of heptazine-based chromophores with hydroxylic partners. Like water molecules, phenol molecules form a hydrogen bond with the peripheral nitrogen atoms of Hz or TAHz. The effect of hydrogen bonding can be observed in the absorption spectrum of TAHz, as is shown in Fig. 9. Addition of 100 mM PhOH to TAHz in toluene causes a distinct red-shift of the absorption threshold of TAHz near 400 nm.[50] On the other hand, weak peaks arising from n$\pi$* states are blue-shifted due to hydrogen bonding and gain intensity, see the inset in Fig. 9. From the shift of the absorption spectrum as a function of PhOH concentration, the association constant of TAHz and PhOH can be determined.[50]

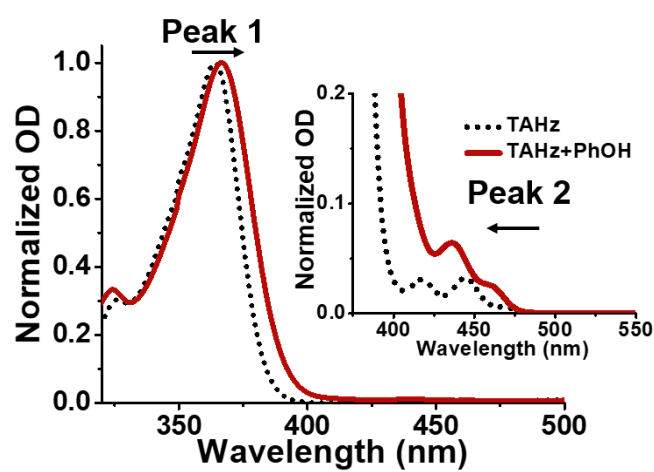

*Fig. 9. The addition of phenol (PhOH) changes the ground-state absorption of TAHz in toluene, suggesting hydrogen-bonding. A distinct redshift of the main absorption can be seen from TAHz in toluene without PhOH present (black dotted line) upon the addition of 100 mM PhOH (solid red line). The weakly allowed transitions at lower energies, on the other hand, exhibit a blueshift upon addition of PhOH, shown in the inset. [Adapted with permission from E. J. Rabe et al., J. Phys. Chem. C **123**, 29580 (2019). Copyright American Chemical Society 2019].*

The reactivity of phenol in hydrogen-bonded complexes can be tuned by functional groups R, for example in the *para* position. It has been shown that the oxidation potential of phenols is shifted cathodically (anodically) by electron donating (electron withdrawing) substituents.[22-26] To demonstrate the possibility of chemical control of the PCET reaction in TAHZ···phenol complexes, we considered the series of complexes TAHz···R-PhOH with R = MeO, Me, H, Br, Cl, CN. These groups range from strongly electron donating (MeO) to strongly electron withdrawing (CN). The kinetic studies were conducted for the TAHz···R-PhOH complexes in toluene solution. The accompanying ab initio calculations were performed, for reasons of computational feasibility and cost, for Hz···R-PhOH complexes in the gas phase. The calculated Hz···R-PhOH hydrogen-bond lengths range from 1.904 Å for R = CN to 1.943 Å for R = MeO. Electron donating substituents increase the electron density on the hydroxyl group and thereby reduce the strength of hydrogen bonding.

The calculated excitation energies and oscillator strengths of the lowest three locally excited $^1\pi\pi^*$ states and of the lowest intermolecular CT state of the six Hz-phenol complexes are listed in Table 3. The dark $^1n\pi^*$ states are omitted for clarity. The excitation energies of the locally excited states of Hz are little affected by the substituent R on PhOH (with the exception of MeO). The vertical excitation energy of the CT state (marked in bold in Table 3), on the other hand, is highly sensitive to R. It drops by more than one electron volt from 3.86 eV for R = CN to 2.80 eV for R = MeO.

**Table 3.** Vertical excitation energies (in eV) of the lowest three $^1\pi\pi^*$ excited states of the Hz core and of the intermolecular CT state of Hz-R-PhOH complexes at their ground-state equilibrium geometries. The excitation energy of the CT state is marked in bold. The numbering of the $^1\pi\pi^*$ states varies for different substituents R due to the interspersed $^1n\pi^*$ states.

| R | CN | Cl | Br | H | CH$_3$ | OCH$_3$ |
|---|---|---|---|---|---|---|
| S$_1$($\pi\pi^*$) | 2.61 | 2.60 | 2.60 | 2.60 | 2.60 | 2.59 |
| S$_n$($\pi\pi^*$) | **3.86** | **3.45** | **3.44** | **3.44** | **3.23** | **2.80** |
| S$_n$($\pi\pi^*$) | 4.40 | 4.39 | 4.39 | 4.39 | 4.25 | 3.83 |
| S$_n$($\pi\pi^*$) | 4.45 | 4.44 | 4.43 | 4.33 | 4.32 | 3.89 |

The pronounced variation of the energy of the intermolecular CT state is reflected by a corresponding strong variation of the topography of the relaxed PE surface for PCET in Hz···R-PhOH complexes. These surfaces were computed for R = CN, H, and MeO and are displayed in Fig. 10(a)-(c). The circle marks the energy minimum of the S$_1$($\pi\pi^*$) state and the triangle marks the saddle point for the PCET reaction. It can be seen that the reaction barrier decreases from R = CN to R = H and is absent for R = MeO. The decrease of the barrier height from R = CN to R = MeO is correlated with the excitation energy of the CT state. The variation of the substituent R thus allows a systematic tuning of the excited-state PCET reaction from a barrier-controlled reaction in Hz···CN-PhOH (E$^\ddagger$ = 0.37 eV) via a low-barrier reaction in Hz···PhOH (E$^\ddagger$ = 0.17 eV) to a barrierless reaction in Hz···MeO-PhOH. It should also be noted that the PCET barrier for Hz···PhOH ($E^‡ = 0.17$ eV) is considerably lower than the barrier for the Hz···$H_2O$ complex ($E^‡ = 0.98$ eV) which reflects the comparatively low O-H bond dissociation energy of phenol[51] ($D_0 = 3.72$ eV, 1.38 eV lower than that of $H_2O$).

PCET rate constants and KIEs (PhOH/D) were measured for TAHz···R-PhOH complexes[50] and more details can be found in this publication. The TAHz···CN-PhOH complex exhibits the lowest PCET rate. The rate measured for the TAHz···Me-PhOH complex is faster by a factor of about ten. For the TAHz···MeO-PhOH complex, the increase of the rate appears to saturate, which may indicate that other factors, such as the association constant, may become rate determining. As shown in Fig. 11, the rates measured for the TAHz···R-PhOH complexes exhibit a reasonable correlation with the barriers computed for the Hz···R-PhOH complexes.

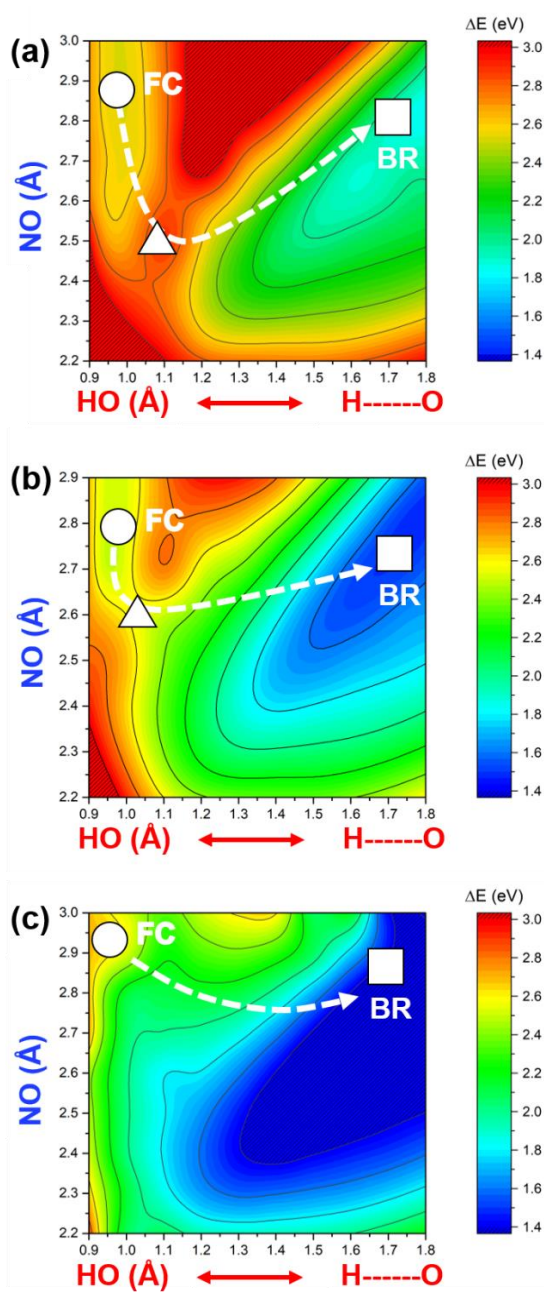

*Fig. 10.* 2D relaxed potential energy surface of the $S_1$ excited state of the Hz···CN-PhOH complex (a), the Hz···PhOH complex (b), and the Hz···$OCH_3$-PhOH complex (c) computed with the ADC(2) method. [Adapted with permission from E. J. Rabe et al., J. Phys. Chem. C **123**, 29580 (2019). Copyright American Chemical Society 2019].

It should be noted that the measured rate of excited-state PCET in TAHz···R-PhOH complexes is anti-correlated with the strength of the hydrogen bond in the electronic ground state. The hydrogen-bond strength increases from R = MeO to R = CN, while the excited-state PCET rate decreases along the series. While a higher association constant due to a stronger hydrogen bond favors PCET, it obviously is not the dominant factor. This illustrates once more that the decisive factor for the rate of the PCET reaction is the energy of the intermolecular CT state which decreases substantially from R = CN to R = MeO.

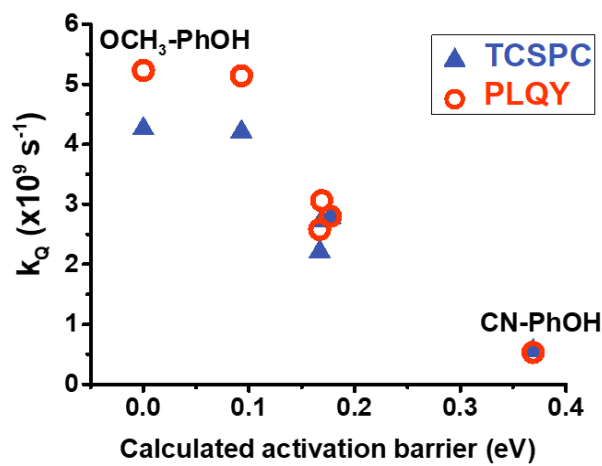

*Fig. 11.* Correlation between the calculated activation barrier and the excited-state quenching rate constants, determined by the photoluminescence quantum yield (PLQY) (red open circles) and by time-correlated single-photon counting (TCSPC) (blue triangles). The excited-state quenching rate increases significantly as the phenol substituent becomes more electron-donating. [Adapted with permission from E. J. Rabe et al., J. Phys. Chem. C **123**, 29580 (2019). Copyright American Chemical Society 2019].

### 3.3. Complexes of functionalized heptazines with water

The electronic-structure calculations and kinetic measurements on Hz-phenol complexes revealed that substituents on phenol can have a dramatic effect on the excited-state energy landscape of these complexes. In particular, the excitation energy of the intermolecular CT state and, as a consequence, the PCET barrier were found to be highly sensitive to substituents on phenol. These findings suggest that the PCET reactivity in complexes of Hz with hydroxylic substrate molecules may alternatively be manipulated by substituents on Hz. The effect of a given substituent on the chromophore (the electron acceptor) should be opposite to the effect of the same substituent on the substrate molecule (the electron donor). Therefore, electron withdrawing groups like CN should lower the intermolecular CT state and the PCET barrier, while electron donating substituents like MeO or anisole should raise the CT state and the PCET barrier. Because a low PCET barrier favors H-atom abstraction from a solvent molecule, electron-withdrawing groups should enhance the efficiency of Hz-based water oxidation photocatalysts.

Because the synthesis, purification and structural characterization of tailored Hz derivatives is difficult and time consuming, guidance by computational studies can help in the search for optimally functionalized chromophores. We therefore performed a comparative computational study of complexes of three tri-substituted heptazines with $H_2O$. In the order of decreasing electron withdrawing character of the substituent, we included tri-cyano-heptazine (TCNHz), tri-chloro-heptazine (TClHz), Hz, and TAHz in the comparison.[52]

The vertical excitation energies of the three lowest locally excited $^1\pi\pi$ states and of the lowest intermolecular CT state of the four chromophore-water complexes are listed in Table 4. The oscillator strengths are given in parentheses. The dark $^1n\pi^*$ states are omitted for clarity. The "small" substituents Cl and CN cause only moderate shifts (of the order of 0.3 eV) of the locally excited states of the Hz core. The pronounced effect of the anisole substituents of the excitation energy of the bright $^1\pi\pi^*$ state of Hz, resulting in a red-shift of 0.8 eV, was discussed in Section 2.2. The effect of the substituents on the energy of the CT state is remarkable. From 5.40 eV in the Hz$\cdots$H$_2$O complex, the CT state is red-shifted to 4.13 eV in the TCNHz$\cdots$H$_2$O complex. On the other hand, it is blue-shifted to 6.15 eV in the TAHz$\cdots$H$_2$O complex. This variation of the vertical excitation energy of the CT state by two electron volts clearly must have pronounced consequences for the photoreactivity of these complexes.

**Table 4.** Vertical excitation energies (in eV) of the lowest three locally excited $^1\pi\pi^*$ states and the lowest CT state of hydrogen-bonded complexes of Hz and three Hz derivatives with a water molecule, calculated with the ADC(2) method. Oscillator strengths are given in parentheses. The charge-transfer state is marked in bold.

| Hz$\cdots$H$_2$O | TClHz$\cdots$H$_2$O | TCNHz$\cdots$H$_2$O | TAHz$\cdots$H$_2$O |
|---|---|---|---|
| $S_1(\pi\pi^*)$  2.59 (0.000) | $S_1(\pi\pi^*)$  2.79 (0.000) | $S_1(\pi\pi^*)$  2.33 (0.000) | $S_1(\pi\pi^*)$  2.65 (0.001) |
| $S_5(\pi\pi^*)$  4.43 (0.258) | $S_5(\pi\pi^*)$  4.48 (0.325) | **$S_5$(CT)  4.13 (0.000)** | $S_2(\pi\pi^*)$  3.50 (1.083) |
| $S_6(\pi\pi^*)$  4.43 (0.283) | $S_6(\pi\pi^*)$  4.48 (0.296) | $S_6(\pi\pi^*)$  4.16 (0.311) | $S_3(\pi\pi^*)$  3.59 (1.208) |
| **$S_n$(CT)  5.40 (0.002)** | **$S_8$(CT)  4.96 (0.002)** | $S_7(\pi\pi^*)$  4.17 (0.287) | **$S_n$(CT)  6.15 (0.037)** |

In Fig. 12, the relaxed $S_1$ PE surfaces for the PCET reaction in the TCNHz$\cdots$H$_2$O, TClHz$\cdots$H$_2$O and TAHz$\cdots$H$_2$O complexes are compared with the PE surface of the Hz$\cdots$H$_2$O complex which has been discussed above (Section 3.1). For all systems, the adiabatic $S_1$ PE surface as function of the H-atom transfer coordinate $R_{OH}$ and the donor-acceptor distance $R_{ON}$ exhibits two minima, one corresponding to the lowest locally excited $^1\pi\pi^*$ state of the chromophore-H$_2$O hydrogen-bonded complex, the other to the biradical in which an H-atom has been transferred from the H$_2$O molecule to the chromophore. The energies of the locally excited $S_1$ minima are stabilized by less than 0.1 eV with respect to the vertical excitation energies, confirming very weak electronic-vibrational coupling in the locally excited $S_1$ state of all complexes. The barriers separating the two minima, on the other hand, are significantly modified by the substituents on the Hz core. The electron-donating group anisole increases the barrier from 0.98 eV for Hz$\cdots$H$_2$O to 1.08 eV for TAHz$\cdots$H$_2$O. The strongly electron withdrawing substituent CN, on the other hand, lowers the barrier to merely 0.59 eV. The PCET barrier thus can be manipulated by 0.5 eV by varying the functional group on Hz. Since the tunnelling rate depends exponentially on the height of the barrier (assuming constant width), this finding implies substantial tunability of PCET rates by functionalization of the Hz chromophore. In general, the efficiency of the photooxidation of hydroxylic solvent molecules should be enhanced by electron withdrawing groups on Hz.

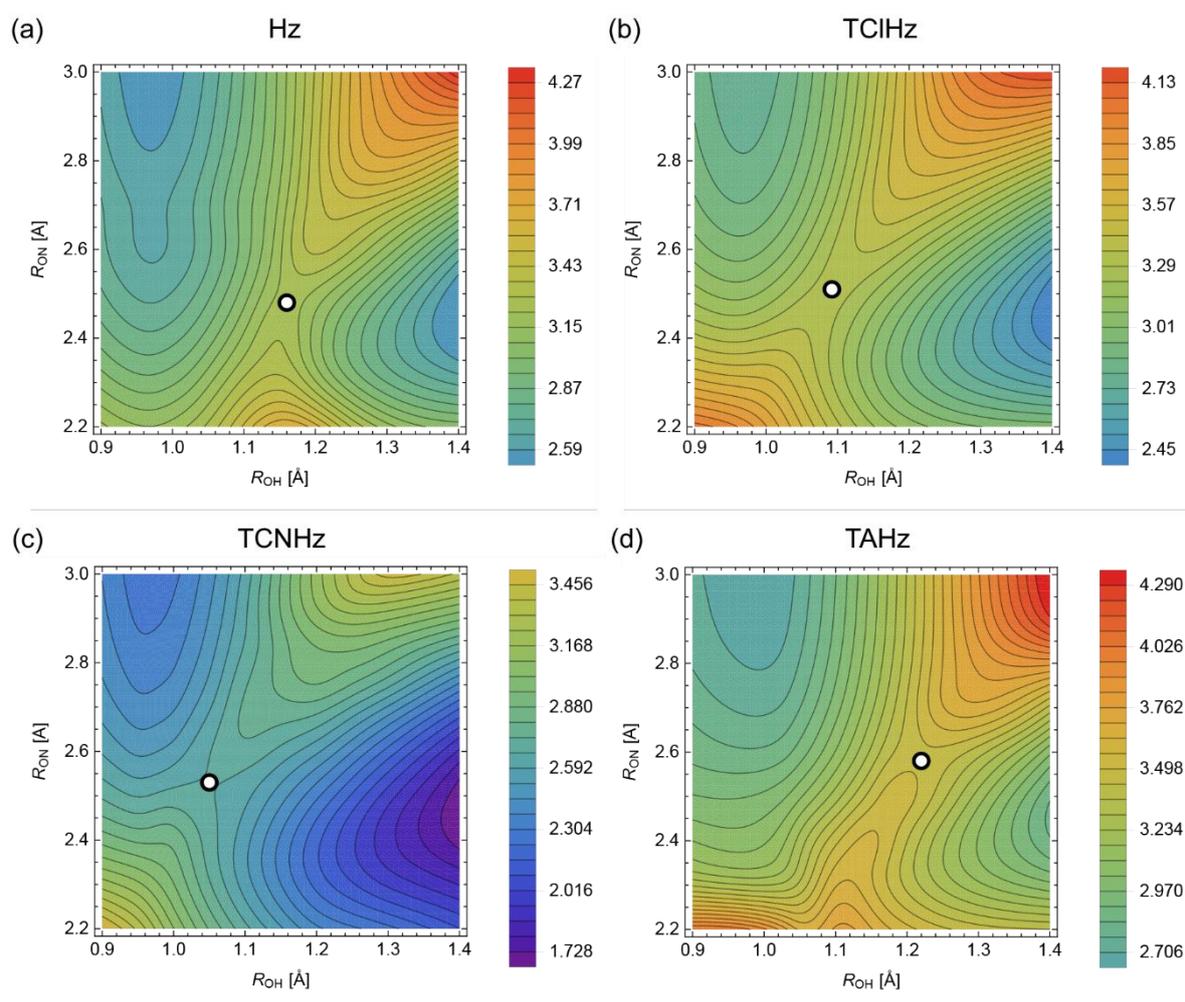

*Fig. 12.* Two-dimensional relaxed PE surfaces of the $S_1$ state of the (a) Hz···$H_2O$, (b) TClHz···$H_2O$, (c) TCNHz···$H_2O$ and (d) TAHz.···$H_2O$ complexes. The saddle point is marked by the circle. Energies are given in eV relative to the energy minimum of the $S_0$ state. The height of the barrier relative to the energy minimum of the locally excited $S_1$ state is 0.98 eV in (a), 0.80 eV in (b), 0.59 eV in (c), and 1.08 eV in (d). [Adapted with permission from J. Ehrmaier et al., J. Phys. Chem. A **124**, 3698 (2020). Copyright American Chemical Society 2020].

It should be kept in mind that the yield of biradicals is also determined by the nonadiabatic dynamics at the seam of intersection of the PE surface of the biradical with the PE surface of the closed-shell state, as has been discussed in Section 3.1. At this intersection seam, the photochemical dynamics bifurcates into the formation of the radical pair (successful H-atom transfer) and the formation of the closed-shell state. The latter outcome implies relaxation back to the ground state of the chromophore-water complex (aborted H-atom transfer). Nonadiabatic electronic/nuclear dynamics simulations of the type recently performed for pyridine-$(H_2O)_n$,[53] pyrimidine-$(H_2O)_n$[54] or Hz-$(H_2O)_n$[55] complexes can provide deeper insight how nonadiabatic transitions and H-atom tunnelling dynamics determine the branching ratios of the photochemistry.

### 3.4. Laser control of PCET in TAHz-phenol complexes

As discussed in Section 3.2, hydrogen-bonded complexes of TAHz with functionalized phenols, R-PhOH, provide versatile models for the investigation of excited-state PCET under precisely defined conditions. In particular, the systematic tuning of the reaction barrier by the substituent R on phenol allows the investigation of the PCET reaction for a range of scenarios, from the high-barrier via the low-barrier to the barrierless case. While passive control of photochemistry by chemical tuning of the reactants has been demonstrated previously (see, e.g.,[24-26]), scenarios of active laser control of PCET photochemistry have remained unexplored so far. In recent work, Corp et al. demonstrated the possibility of active steering of PCET reactivity with so-called pump-push-probe spectroscopy for TAHz⋯R-PhOH complexes.[56]

Pump-push-probe spectroscopy is a multi-pulse ultrafast transient absorption technique which recently was employed by a number of research groups for the monitoring of charge-separation dynamics in organic semiconductors.[57-60] An actinic pump pulse, which generates an impulsive electronic excitation, is followed after a certain time delay by an impulsive push pulse which instantaneously imparts additional energy to the system. By this additional energy, the system may overcome activation barriers to access otherwise inaccessible reaction channels. The reactivity is monitored by the probe pulse via transient absorption. In the context of photochemical reactivity of molecular systems, pump-push-probe spectroscopy depends on the existence of a long-lived excited state which provides a reservoir of excited-state electronic population from which higher excited electronic states can be selectively excited in a time and frequency resolved manner. If the higher excited states are nonreactive, the electronic population relaxes back to the long-lived reservoir state on timescales of less than 200 fs and no persistent loss of the transient absorption signal ($\Delta\Delta$OD) is observed on picosecond timescales. If, on the other hand, the upper excited states populated by the push pulse are reactive, the excited-state reservoir is depopulated by the push pulse which is detected as a persistent quenching of the transient absorption signal.

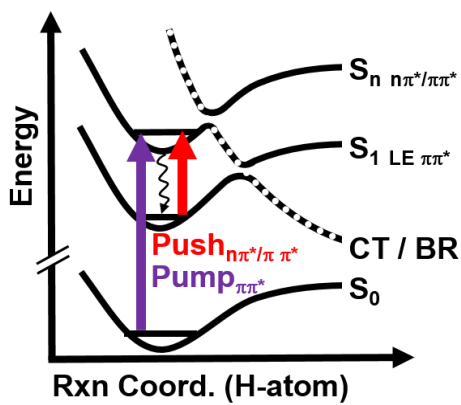

*Fig. 13. Cartoon potential energy diagram showing a photochemical proton transfer reaction driven by an electronic charge transfer (CT) state which is strongly stabilized by the proton transfer, resulting in the formation of a biradical (BR). The push pulse populates higher excited states ($S_n$) which can efficiently couple to the CT state. [Adapted with permission from K. L. Corp et al., J. Phys. Chem. C **124**, 9151 (2020). Copyright American Chemical Society 2020].*

A schematic illustration of the principle of this experiment for Hz-solvent complexes is illustrated in Fig. 13. The excited state populated by the pump pulse is the bright $^1\pi\pi^*$ state of the Hz-based chromophore, which may be mixed with dark $n\pi^*$ states and is designated as $S_n(\pi\pi^*/n\pi^*)$ in Fig. 13. The reservoir state is the unique long-lived lowest $^1\pi\pi^*$ state ($S_1$) of Hz. As indicated in Fig. 13, the interaction of the locally excited states with the reactive CT state generates barriers for PCET such that the barrier in the upper $S_n(\pi\pi^*/n\pi^*)$ states is much lower than the barrier in the $S_1$ state. As a result, excitation of the $S_1$ population by the push pulse (red arrow in Fig. 13) leads to rapid PCET which can be observed by the persistent loss of chromophores in the $S_1$ state.

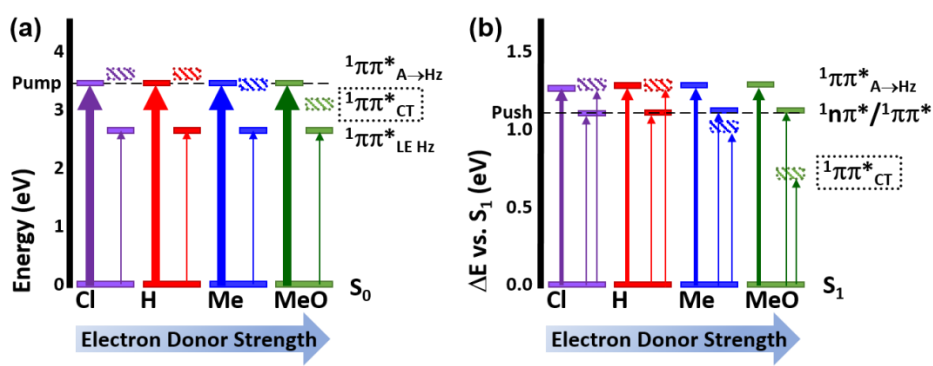

*Fig. 14. (a) Diagram depicting transition energies from the electronic ground state to the locally excited $^1\pi\pi^*$ state and the CT state for four TAHz⋯R-PhOH complexes (R = Cl, H, Me, MeO). Calculated oscillator strengths of transitions are represented by the line width of the vertical arrows. (b) Diagram depicting transition energies from the relaxed geometry of the $S_1$ state. Pump and push energies are represented by the black horizontal dashed lines. Note the different energy scales in (a) and (b). [Adapted with permission from K. L. Corp et al., J. Phys. Chem. C **124**, 9151 (2020). Copyright American Chemical Society 2020].*

The pump-push-probe measurements were performed for four selected TAHz⋯R-PhOH complexes, R = Cl, H, Me, MeO. The experiments were guided by ab initio calculations of transition energies and transition strengths. Fig. 14(a) shows the calculated transition energies from the electronic ground state to the two locally excited $^1\pi\pi^*$ states and the intermolecular CT state for the four TAHz⋯R-PhOH complexes in graphical form. As discussed in the preceding section, the energies of the locally excited states are insensitive to the substituent R, while the energy of the CT state is significantly lowered by electron donating groups (Me, MeO). Fig.14(b) gives the corresponding information for the electronic transitions from the minimum of the long-lived $S_1$ state. Again, the excitation energies of the locally excited states are insensitive to R, whereas the excitation energy of the CT state is strongly reduced by the electron withdrawing groups Me and MeO. Based on these data, the push pulse was tuned to 1150 nm (1.08 eV) in the experiment.

Fig. 15(a) shows the transient absorption spectra of TAHz in neat toluene ("No PhOH") in comparison with the spectra of TAHz in the presence of PhOH, Cl-PhOH, Me-PhOH, and MeO-PhOH. The solution was pumped at 365 nm, pushed at 1150 nm and the transient absorption kinetics detected at 700-750 nm.[56] The pump-push delay was 6 ps. As shown in Fig. 15(b), a persistent quenching of the transient absorption signal by the push pulse is observed when PhOH is present. The magnitude of this quenching decreases with the electron withdrawing character of the substituent on PhOH and is non-existent for R = MeO. This result is in full agreement with the conclusions drawn in Section 3.2, based on measurements for TAHz···R-PhOH complexes and calculations for Hz···R-PhOH complexes. While a low barrier exists for PCET in the $S_1$ state of TAHz···PhOH and TAHz···Cl-PhOH, this barrier becomes tiny for TAHz···Me-PhOH and is absent for TAHz···MeO-PhOH. For the latter complex, the PCET reaction occurs spontaneously in the $S_1$ state and therefore no long-lived reservoir of excited electronic population exists. For TAHz···PhOH and TAHz···Cl-PhOH, on the other hand, the pump-push-probe experiment reveals strong enhancement of the PCET reaction in the $S_n(\pi\pi^*/n\pi^*)$ excited states due to their quasi-degeneracy with the reactive CT state.

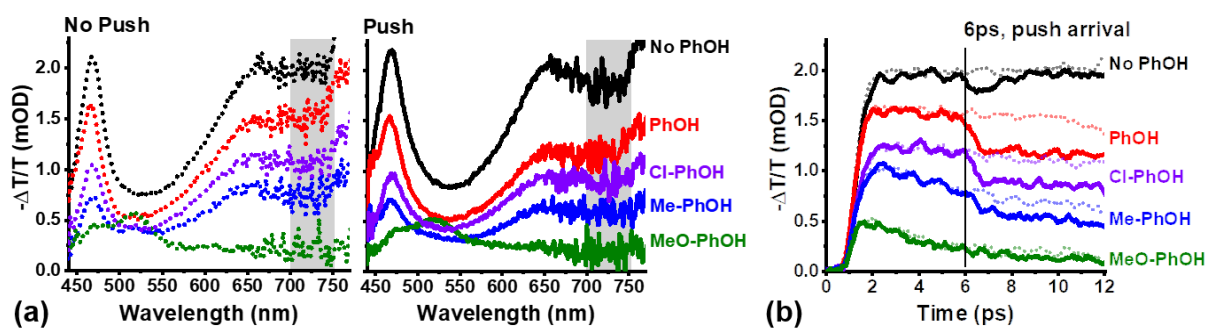

*Fig. 15. (a) Left panel, dotted lines: Pump-probe spectra of TAHz (50 μM) in toluene with and without phenol derivatives, R-PhOH. Right panel, solid lines: Pump-push-probe spectra of TAHz (50 μM) in toluene with and without phenol derivatives, R-PhOH. (b) Decay of the population probed from 700 – 750 nm (shaded in gray in (a)) with (solid) and without (dotted) the push pulse. The system was pumped at 365 nm and pushed at 1150 nm at 6 ps. [Adapted with permission from K. L. Corp et al., J. Phys. Chem. C **124**, 9151 (2020). Copyright American Chemical Society 2020].*

The rational laser control of PCET dynamics guided by ab initio calculations of the energy landscape of excited electronic states provides unprecedentedly detailed insight into the mechanisms of the H-atom transfer photochemistry in Hz-solvent complexes, in particular into the branching mechanisms among unreactive locally excited states and reactive intermolecular CT states. Based on this insight, knowledge-driven design strategies for efficient water oxidation photocatalysts can be developed in the future which should prove useful for synthetic chemists and materials scientists.

## 4. Mechanism of self-repair of Hz-based photocatalysts

While the decomposition of $H_2O$ molecules into H and OH radicals with visible or near infrared light has been a long-standing challenge in photochemistry, an even bigger challenge is the taming of the *in situ* generated OH radicals. OH radicals are highly reactive and exhibit an exceptionally high diffusion rate. To avoid or at least reduce detrimental reactions of OH radicals with (photo)catalysts, the OH radicals generated by hydrogen abstraction from water are scavenged in current experiments on molecular hydrogen evolution by sacrificial hole scavengers. In the future, the OH radicals should be captured and recombined in a controlled manner to the closed-shell products $H_2O_2$ or $H_2O$ plus $O_2$ by suitable catalysts.

Once recombination with OH radicals is suppressed, the photogenerated reduced Hz radicals can be harvested and recombined to form $H_2$, which regenerates the photocatalyst. The most likely undesired side reaction of photochemically generated radical pairs is *in situ* geminate radical recombination, in the present case the recombination of HzH and OH radicals. In particular, the CH group in the immediate vicinity of the reduced N-atom in the HzH radical is prone to be attacked by OH radicals, because the CN bond is weakened by the reduction of the N-atom. Computational studies have shown that the OH radical can attach to this CH group, forming a covalent bond.[61] The structure of the resulting closed-shell product, which formally is a hydrate of Hz, is shown in Fig. 16(b). A moderately endothermic ring-opening reaction can lead to the hydroxy-imino isomer displayed in Fig. 16(c). An exothermic reaction finally leads to the completely open structure shown in Fig. 16(d), see also Scheme 1(b). This photohydrate of Hz is predicted to be more stable than the original Hz⋯$H_2O$ complex by about 3 kcal/mol at the MP2/cc-pVDZ level.

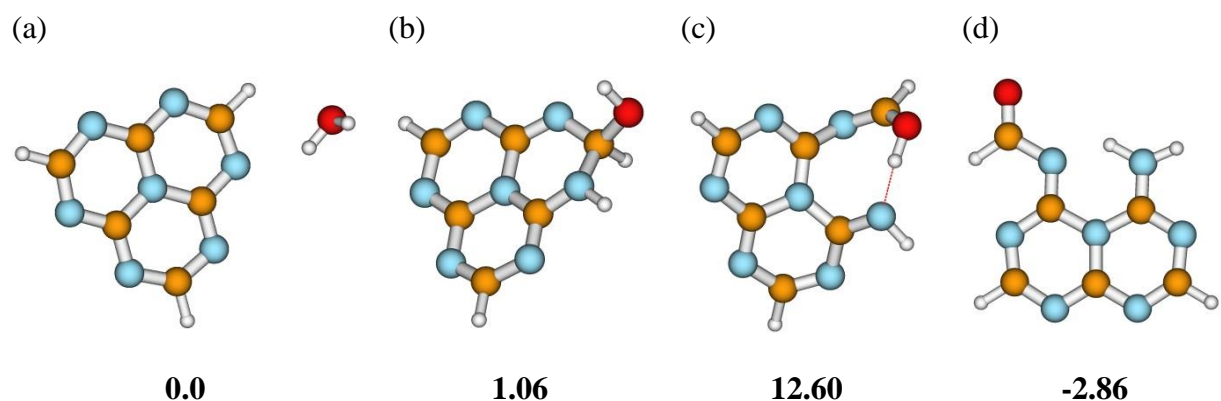

*Fig. 16. Structures and relative energies (in kcal/mol) of the Hz⋯$H_2O$ complex (a) and of photohydrates of Hz (b - d), computed at the MP2/cc-pVDZ level. [Adapted with permission from W. Domcke et al., ChemPhysChem **3**, 10 (2019). Copyright Wiley VCH 2019].*

The existence of a thermochemically stable hydrate of Hz which is accessible by a low-barrier photochemical reaction explains the irreversible hydrolysis of the Hz molecule in the presence of water and light.[31, 32] The predicted lowest excitation energy of the photohydrate is 3.5 eV (355 nm) at the ADC(2)/cc-pVDZ level.[61] The photohydrate therefore is colorless, which excludes the back-conversion of the photohydrate to Hz and water under visible light.

Table 5 lists the energy of the most stable form of the photohydrate relative to the energy of the chromophore-water complex for a series of substituted heptazines. A negative entry indicates thermochemical stability of the photohydrate, a positive entry indicates metastability. The data exhibit a clear trend: electron withdrawing substituents on Hz tend to stabilize the photohydrate, while electron donating substituents tend to destabilize the photohydrate. Anisole is an electron donating group and the photohydrate of TAHz indeed is found to be metastable by 5.85 kcal/mol, see Table 5. It therefore can convert back to TAHz and $H_2O$ by a thermal reaction in the dark.

**Table 5.** Energies (in kcal/mol) of the intermediate photohydrate (structure (b) in Fig. 13) and the thermodynamically most stable photohydrate (structure (d) in Fig. 13) relative to the energy of the chromophore-water complex for Hz···$H_2O$, TClHz···$H_2O$, TCNHz···$H_2O$ and TAHz···$H_2O$, calculated at the MP2/cc-pVDZ level.

| substituent | photohydrate (b) | photohydrate (d) |
| --- | --- | --- |
| H | 1.06 | -2.86 |
| Cl | -5.98 | -5.32 |
| CN | -3.56 | -5.88 |
| anisole | 11.33 | 5.85 |

The prediction that TAHz has self-repairing capability is in agreement with the observations in the laboratory. It has been found that TAHz exhibits excellent long-term stability under UV irradiation, in sharp contrast to Hz. In $H_2$ evolution experiments driven at 365 nm in an aqueous suspension of TAHz with 2% Pt loading and TEOA as sacrificial electron donor, the $H_2$ evolution did not decrease over 25 days of continuous illumination.[21] The photostability of TAHz is comparable to that of polymeric carbon nitride materials.[1, 5] It is remarkable that the celebrated *operando* stability of polymeric carbon nitrides can be matched by a molecular chromophore in solution/suspension. It is likely that self-repair after radical recombination reactions plays an essential role for the exceptional durability of molecular TAHz in aqueous environments under UV irradiation.

## 5. Closing the photocatalytic cycle

The products of the photochemical hydrogen abstraction reaction from protic substrates by Hz-based chromophores are substrate-derived radicals (e.g, hydroxyl or phenoxyl radicals) and reduced chromophore radicals (e.g., HzH or TAHzH). The substrate-derived radicals are waste products and ideally should be recombined to stable closed-shell products (e.g, 2 ·OH → $H_2O_2$). The reduced chromophore, on the other hand, carries part of the energy of the absorbed photon as chemical energy. To render the photochemical H-atom abstraction reaction catalytic, the chromophore has to be recovered and the chemical energy stored in the reduced chromophore radical has to be converted to a stable and useful chemical product. Here, we briefly discuss two scenarios for the closure of the photocatalytic cycle.

The HzH radical is a chemically stable species. Its dissociation energy has been estimated as 2.0 eV (46 kcal/mol).[20] Since the dissociation energy $D_0$ of $H_2$ is 4.48 eV,[62] the recombination of two HzH radicals, forming $H_2$ and regenerating two Hz molecules, is exothermic. In the first scenario, the catalytic cycle is closed by the recombination of two HzH radicals in a thermal reaction. Overall, two photons are absorbed by two different Hz chromophores and (assuming water as the substrate) two $H_2O$ molecules are thereby decomposed into $H_2$ and two OH radicals. The evolution of $H_2$ with molecular TAHz as photocatalyst and with TEOA as OH radical scavenger[21] likely proceeds via this mechanism. The nano-scale Pt co-catalyst may facilitate the recombination of the TAHzH radicals.

In the second scenario, the chromophore, say Hz, is recovered by photodetachment of the excess H-atom from the HzH radical, see Scheme 1(c). It has been shown that the HzH radical as well as substituted HzH radicals possess so-called πσ* excited states which play an essential role for their photochemistry.[20, 52] The characteristic photochemistry of πσ* excited states is well established for protic aromatic chromophores such as indole, aniline or phenol.[63-67] The $^1$πσ* states in these systems are repulsive along the N-H or O-H stretching coordinate. The energy of the dark dissociative $^1$πσ* state typically crosses the energy of the bright $^1$ππ* state as well as the energy of the $S_0$ state. The associated conical intersections provide an efficient mechanism for radiationless deactivation of the $^1$ππ* state via H-atom photodetachment or internal conversion to the $S_0$ state.[68] These photoreactions were comprehensively characterized by energy-resolved as well as time-resolved detection of the free H-atoms.[63, 64, 66, 67]

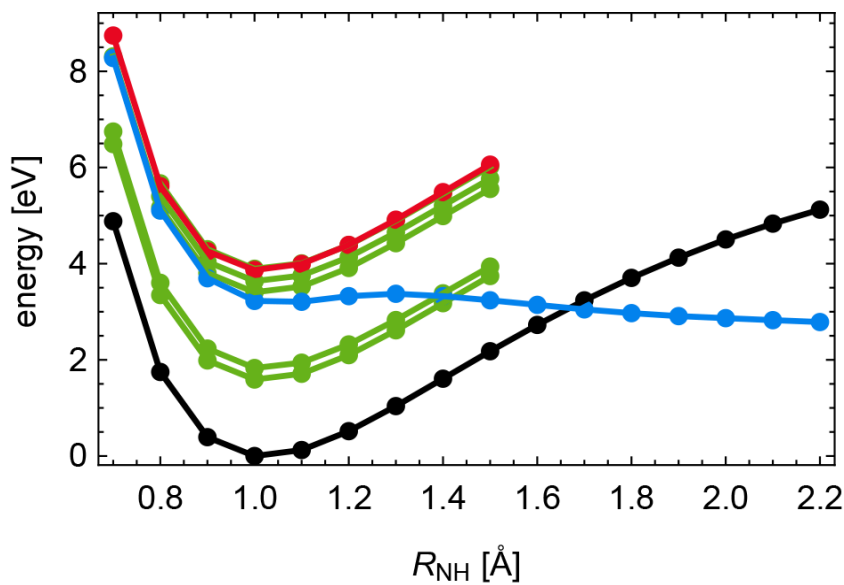

*Fig. 17. Energy profiles of the electronic ground state (black) and the lowest excited electronic states along the NH stretching coordinate of the heptazinyl radical, calculated with the ADC(2) method. Green: $^2\pi\pi^*$ excited states; red: $^2n\pi^*$ state; blue: $^2\pi\sigma^*$ state. [Adapted with permission from J. Ehrmaier et al., J. Phys. Chem. A **121**, 4754 (2017). Copyright American Chemical Society 2017].*

The vertical electronic excitation spectra of the HzH radical and a few HzH derivatives were computed with the unrestricted ADC(2) method.[69] As an open-shell radical, HzH possesses a dense spectrum of low-energy excited electronic states. The PE profiles of the lowest electronic states of the HzH radical as a function of the N-H stretching coordinate are displayed in Fig. 17. The PE functions of the $^2\pi\pi^*$ excited states (shown in green) and of the $^2n\pi^*$ excited states (shown in red) are parallel to the PE function of the electronic ground state (shown in black). On the other hand, the PE function of the lowest $^2\pi\sigma^*$ state (shown in blue) is dissociative with respect to N-H bond stretching of HzH. The $^2\pi\sigma^*$ state is the third excited state at the ground-state equilibrium geometry of HzH and its repulsive PE function crosses the PE functions of the lowest two $^2\pi\pi^*$ states as well as the PE function of the electronic ground state. These allowed curve crossings become symmetry-allowed conical intersections when out-of-plane vibrational modes are taken into account. The nonadiabatic dynamics at the conical intersections determines the branching ratio of direct (nonstatistical) H-atom photodetachment vs. radiationless relaxation to the electronic ground state of HzH (aborted photodetachment). Ab initio nonadiabatic dynamics simulations are required for estimates of the photodissociation probability.

While the thermal recombination of HzH radicals (Scenario 1) yields molecular hydrogen, the photodetachment reaction (Scenario 2) in the gas phase or in nonpolar solution yields free hydrogen atoms. When the photodetachment takes place in aqueous solution, on the other hand, $H_3O$ radicals are intermediately formed which relax to the so-called hydrated electron by a spontaneous charge separation process.[70] Recombination of these H-atoms or hydrated electrons to $H_2$ is undesirable, since H-atoms or hydrated electrons are more powerful reducing agents than molecular hydrogen.

There should be a wide range of opportunities for the production of high-value chemicals from cheap precursors by exploitation of the exceptional reduction potential of atomic hydrogen or hydrated electrons.

## 6. Discussion

At the dawn of the research field of artificial photosynthesis it has been argued that radical intermediates have to be avoided in water-splitting scenarios owing to their high energy and high reactivity.[71] The research field has followed this paradigm since the 1970s. The overwhelming part of the literature on photocatalytic water splitting is concerned with two-electron (e.g., $2H^+ + 2e^- \rightarrow H_2$) or four-electron (e.g., $2H_2O \rightarrow 2H_2 + O_2$) redox reactions between closed-shell species excluding radical intermediates.[72-75] However, despite decades of intensive research, an unequivocal demonstration that strongly endothermic multi-electron redox reactions can be driven purely by photons (without additional support by overpotentials in electrochemical devices) is lacking. It is now widely recognized that the bottleneck in water-splitting photocatalysis is the four-electron redox reaction which is required for the evolution of $O_2$.[76-80]

Photochemical redox reactions driven by low-intensity irradiation invariably are one-electron transfer reactions because photons couple to single electrons and electron-electron correlation in atoms and molecules is too weak to enable multi-electron processes driven by a single photon. One-electron transfer reactions starting from closed-shell educts only can yield radical intermediates. The typical rate for sequential absorption of photons under solar irradiation is $10^{-3}$ $s^{-1}$, which is slow compared to characteristic rates of photoinduced intramolecular processes as well as molecular diffusion and intermolecular reaction processes. Therefore, multi-electron redox reactions can realistically only be realized as sequential absorption processes of several photons. This calls for a reconsideration of the strict dogma of exclusion of radical intermediates in water oxidation photocatalysis.

The argument against the intermediate generation of radicals[71] ("their production is a sheer waste of energy from the photochemical conversion point of view") overlooks that H and OH radicals are not necessarily produced as free radicals, but may be stabilized by binding to suitable complexation partners. A general model for the photoinduced one-electron homolytic water decomposition reaction can be written as follows

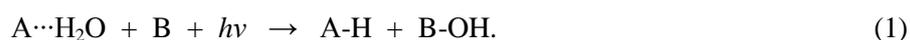

$$A\cdots H_2O + B + h\nu \rightarrow A\text{-}H + B\text{-}OH. \tag{1}$$

Here A is a molecular chromophore (e.g. Hz) which forms a hydrogen bond with an $H_2O$ molecule which is essential for the photoinduced PCET from $H_2O$ to A, as extensively discussed above. B is an (optically inactive) co-catalyst which coordinates and thereby stabilizes OH radicals (examples could be noble-metal, titanium-dioxide, or manganese-oxide nanoclusters). After the sequential light-driven homolytic decomposition of two $H_2O$ molecules according to Eq. (1), the catalytic cycle can be closed by two thermal radical recombination reactions

$$2 \text{ A-H} \rightarrow 2 \text{ A} + H_2 \qquad (2)$$

$$2 \text{ B-OH} \rightarrow 2 \text{ B} + H_2O_2. \qquad (3)$$

In this scenario, $H_2O_2$ is generated as a closed-shell co-product of $H_2$

$$2 H_2O + 2 h\nu \rightarrow H_2 + H_2O_2. \qquad (4)$$

The energetic feasibility of the recombination reactions (2) and (3) requires that the A-H and B-OH bond energies are less than half the bond energies of $H_2$,[62] and $H_2O_2$,[81] respectively, that is,

$$D_0(\text{A-H}) < 2.24 \text{ eV} \qquad (5)$$

$$D_0(\text{B-OH}) < 1.06 \text{ eV}. \qquad (6)$$

Given the dissociation energy of the OH bond of water[82]

$$D_0(\text{H-OH}) = 5.10 \text{ eV}, \qquad (7)$$

two photons of 1.81 eV (687 nm) are required to drive the reaction of Eq. (4). This limiting photon energy of the simple mechanistic scenario of Eqs. (1-3), neglecting solvation effects, is reasonably close to the thermochemical heat of reaction of Eq. (4) (80.3 kcal/mol), which corresponds to two photons of 1.74 eV. While the light-driven generation of $H_2O_2$ has extensively been studied in the literature, for example with titanium dioxide or carbon nitride photocatalysts,[83-86] these investigations were oxygen reduction reactions and should not be confused with the water oxidation reaction of Eq. (4). In reaction (4), $H_2O_2$ is the waste product rather than the primary product.

The advantage of the scenario described by Eqs. (1-4) is the unravelling of the water-splitting problem into simple sequential partial reactions. The thermodynamically most difficult step, the breaking of the OH bond of water, is achieved directly by a light-driven excited-state reaction, which avoids the high barriers on the ground-state PE surface of water. The breaking of the exceptionally strong OH bond of $H_2O$ is facilitated by the formation of the two relatively weak A-H and B-OH bonds. The follow-up reactions (2) and (3), which close the catalytic cycle, are simple exothermic radical-radical recombination reactions. In this scenario, the light-driven chemistry is clearly separated from the thermal follow-up reactions. The photosensitizer (A) and the photo-inactive co-catalyst (B) can separately be optimized. Elusive multi-electron redox reactions have been eliminated.

When the B-OH binding energy is higher than 1.06 eV, the OH radical recombination channel (3) is closed and $H_2O_2$ cannot be formed as a reaction product. In this case, four OH radicals have to be accumulated on the co-catalyst B and have to be recombined to form two $H_2O$ molecules and an $O_2$ molecule. The energy of decomposition of $H_2O_2$ to $H_2O$ and $O_2$ is thus recovered and four photons of 1.23 eV are sufficient to drive the reaction at the thermodynamic limit. While the threshold photon energy of the two-photon scheme of Eq. (4) (1.74 eV) is higher than the threshold photon energy of the standard four-photon water-splitting scheme (1.23 eV), the scenario of Eqs. (1-3) should nevertheless represent an attractive goal for solar energy harvesting due to its simplicity.

In this perspective article, we focussed on the light-driven part of the reaction and the optimization of the photosensitizer, ignoring the stabilization of OH radicals by co-catalysts. The relevance of pre-existing hydrogen bonds for efficient excited-state PCET reactions has been pointed out previously.[25, 87-92] For the Hz-phenol system in toluene solution, we have found clear spectroscopic evidence for the formation of hydrogen bonds and for the relevance of hydrogen bonding for the intra-complex PCET reaction.[50] The essential role of hydrogen bonding for excited-state PCET is also confirmed by the *ab initio* calculations. On the other hand, the results for functionalized phenols reveal that the strength of the ground-state hydrogen bond is not the decisive factor controlling the PCET reactivity. The most important effect is the modulation of the energy of the intermolecular CT state by electron donating or electron withdrawing substituents on the Hz chromophore or on the substrate. With the exception of several theoretical studies,[93-99] the role of hydrogen-bonding with substrate molecules so far has found little attention in the literature on water oxidation photocatalysis with solid-state carbon nitride materials. It is likely that the phenomena analysed herein for molecular Hz-based chromophores are of similar relevance for the photoreactivity of these materials with protic substrates.

The possibility of easily tuning the absorption threshold by substituents is an important advantage of organic chromophores compared with inorganic semiconductors which exhibit band gaps which cannot be as readily modified. In all Hz-based chromophores, the lowest excited state ($S_1(\pi\pi^*)$) is dark and its energy and properties are rather insensitive with respect to chemical modifications. The second $^1\pi\pi^*$ excited state, $S_n(\pi\pi^*)$, on the other hand, is bright and its excitation energy can readily be tuned by chemical modification with aromatic functional groups. This combination of a very bright and short-lived absorbing state ($S_4$ in Hz) and an exceptionally weakly emitting $S_1$ state, which is efficiently populated on femtosecond timescales from the absorbing state by intramolecular radiationless relaxation, is a very favourable feature for water oxidation photocatalysis. A large absorption cross section is thus combined with an exceptionally long lifetime of the $S_1$ state. The $S_1$-$T_1$ energy inversion additionally suppresses ISC as a quenching channel of the $S_1$ state. As a result, a significant part of the energy of the absorbed photon can be stored in the $S_1$ state of the Hz-$H_2O$ complex for hundreds of nanoseconds, which provides ample opportunity for reactions with the co-catalyst B and the formation of the A-H and B-OH radicals in Eq. (1). The pump-push-probe spectroscopy data in TAHz···R-PhOH complexes (Section 4) are a clear documentation of the role of the $S_1$ state as a long-lived reservoir state in the photophysics of TAHz.

In principle, the energy of the absorption maximum of the bright $S_n(\pi\pi)^*$ state can be lowered by substitution to an extent that it approaches the energy of the $S_1(\pi\pi^*)$ state (typically 2.60 eV). This way, the overlap of the absorption spectrum with the spectral profile of solar radiation can be improved and energy dissipation due to the $S_n(\pi\pi)^* \rightarrow S_1(\pi\pi^*)$ radiationless transition can be reduced. TAHz with an absorption threshold at 400 nm currently appears to be a good compromise for initial mechanistic studies. In future studies, the possibility of lowering of the absorption threshold toward 500 nm or below should be explored.

Perhaps the most valuable data provided by the ab initio electronic-structure calculations are the vertical excitation energies of the dark intermolecular CT state in Hz-substrate complexes and their dependence on functional groups on Hz or on the substrate, because this information is not directly accessible experimentally. The energy of the CT state directly controls the barrier for the H-atom transfer reaction on the $S_1$ PE surface of the complex. The lowering of the energy of the CT state by electron donating groups on phenol reduces the barrier for the PCET reaction and thereby enhances the reaction rate dramatically, which is confirmed by the experimental data. In particular, it has been demonstrated that the PCET barrier can be completely eliminated by strongly electron donating substituents on phenol. The calculations predict that the energy of the intermolecular CT state in Hz-$H_2O$ complexes can likewise be modulated over a wide range by substituents on the Hz core. Electron withdrawing groups on Hz stabilize the energy of the HzH radical, which lowers the energy of the CT state and of the H-atom transfer barrier. In hydrogen-bonded complexes of functionalized heptazines with functionalized protic substrates, the dynamics of PCET reactions can be studied in unprecedented detail due the long-lived population in the $S_1$ state, which can be promoted in a controlled manner by frequency-tuned impulsive push pulses to the reactive region in the vicinity of the PCET barrier.[56]

The ubiquitous photosensitizers in artificial photosynthesis are transition-metal complexes with organic ligands because of their long lifetimes of hundreds of nanoseconds up to microseconds in the lowest triplet state, the archetypical photosensitizer being $Ru(bpy)_3^{2+}$. These long lifetimes, being competitive with molecular diffusion, are essential for productive photochemical transformations. However, these long-lived triplet states are known to readily activate atmospheric molecular oxygen to highly reactive singlet oxygen which in turn destroys organic materials. The resulting lack of durability of organometallic (photo)catalysts under irradiation is one of the major unsolved problems of artificial photosynthesis.[38, 39, 77, 100] The fact that Hz-based chromophores do not possess long-lived triplet states due to robust $S_1/T_1$ energy inversion offers an unexpectedly simple resolution of this long-standing problem. The absence of singlet oxygen activation was confirmed both for molecular Hz-based chromophores (TAHz)[35] as well as for polymeric carbon nitride materials.[40-42] This feature of carbon nitride materials most likely is an important factor contributing to their exceptional *operando* stability.

A second property of carbon nitrides contributing to their high photostability may be the ability of self-regeneration after detrimental radical recombination reactions. Depending on the functionalizing groups on the Hz chromophore, the photohydrates formed by undesired radical recombination reactions can be metastable with respect to the chromophore-water complex, which allows recovery of the chromophore by a spontaneous thermal reaction. Electron donating groups on Hz enhance the capability of self-repair. This interesting feature of Hz-based materials is so far just a theoretical prediction which requires experimental confirmation. It is also predicted that the photohydrates of Hz chromophores exhibit characteristic near-UV absorption bands by which the formation and the decay of the photohydrates could be spectroscopically detected.[52, 61]

OH radicals are meanwhile routinely detected as by-products in photocatalytic hydrogen evolution experiments, for example with g-$C_3N_4$[101, 102] or $TiO_2$[103-107] materials. These findings indicate that the observed reactions actually occur via sequential one-electron redox reactions. This calls for an unbiased consideration of the concept of photochemical water oxidation via sequential photon-driven one-electron redox reactions involving radical intermediates. While the control of the reactivity of the intermediate radicals certainly constitutes a serious challenge, this problem seems to have received little attention so far and possible solutions were not explored. If progress can be made with effective control of radical recombination reactions via new concepts and novel techniques, water splitting by sequential one-electron photoreactions may become a viable alternative to the (so far elusive) concept of solar energy harvesting by multi-electron redox reactions.

**Acknowledgments**

The research reported herein has partially been supported by the Deutsche Forschungsgemeinschaft through the Munich Centre for Advanced Photonics (MAP) and by the U. S. National Science Foundation (NSF) under Grant No. CHE-1846480.

**Data availability statement**

The data that support the findings of this study are available from the corresponding author upon request.

**7. References**

[1] X. Wang, K. Maeda, A. Thomas, K. Takanabe, G. Xin, J. M. Carlsson, K. Domen, and M. Antonietti, Nature Mater. **8**, 76 (2009).
[2] W.-J. Ong, L.-L. Tan, Y. H. Ng, S.-T. Yong, and S.-P. Chai, Chem. Rev. **116**, 7159 (2016).
[3] J. Wen, J. Xie, X. Chen, and X. Li, Appl. Surf. Sci. **391**, 72 (2017).
[4] M. Z. Rahman, K. Davey, and S.-Z. Qiao, J. Mater. Chem. A **6**, 1305 (2018).
[5] Y. Wang, X. Wang, and M. Antonietti, Angew. Chem. Int. Ed. **51**, 68 (2012).
[6] C. Merschjann, T. Tyborski, S. Orthmann, F. Yang, K. Schwarzburg, M. Lublow, M.-C. Lux-Steiner, and T. Schedel-Niedrig, Phys. Rev. B **87**, 205204 (2013).
[7] W. Wei, and T. Jacob, Phys. Rev. B **87**, 085202 (2013).
[8] C. Butchosa, P. Guiglion, and M. A. Zwijnenburg, J. Phys. Chem. C **118**, 24833 (2014).
[9] K. Srinivasu, B. Modak, and S. K. Ghosh, J. Phys. Chem. C **118**, 26479 (2014).
[10] S. Melissen, T. Le Bahers, S. N. Steinmann, and P. Sautet, J. Phys. Chem. C **119**, 25188 (2015).
[11] L. M. Azofra, D. R. MacFarlane, and C. Sun, Phys. Chem. Chem. Phys. **18**, 18507 (2017).

[12] V. W.-h. Lau, I. Moudrakovski, T. Botari, S. Weinberger, M. B. Mesch, V. Duppel, J. Senker, V. Blum, and B. V. Lotsch, Nature Commun. **7**, 12165 (2016).

[13] R. Godin, Y. Wang, M. A. Zwijnenburg, J. Tang, and J. R. Durrant, J. Am. Chem. Soc. **139**, 5216 (2017).

[14] K. L. Corp, and C. W. Schlenker, J. Am. Chem. Soc. **139**, 7904 (2017).

[15] M. Z. Rahman, and C. B. Mullins, Acc. Chem. Res. **52**, 248 (2018).

[16] W. Yang, R. Godin, H. Kasap, B. Moss, Y. Dong, S. A. J. Hillman, L. Steier, R. Reisner, and J. R. Durrant, J. Am. Chem. Soc. **141**, 11219 (2019).

[17] Y. Wang, A. Vogel, M. Sachs, R. S. Sprick, L. Wilbraham, S. J. A. Moniz, R. Godin, M. A. Zwijnenburg, J. R. Durrant, A. I. Cooper, and J. Tang, Nature Energy **4**, 746 (2019).

[18] X. Liu, A. L. Sobolewski, R. Borrelli, and W. Domcke, Phys. Chem. Chem. Phys. **15**, 5957 (2013).

[19] J. Ehrmaier, M. J. Janicki, A. L. Sobolewski, and W. Domcke, Phys. Chem. Chem. Phys. **20**, 14420 (2018).

[20] J. Ehrmaier, T. N. V. Karsili, A. L. Sobolewski, and W. Domcke, J. Phys. Chem. A **121**, 4754 (2017).

[21] E. J. Rabe, K. L. Corp, A. L. Sobolewski, W. Domcke, and C. W. Schlenker, J. Phys. Chem. Lett. **9**, 6257 (2018).

[22] J. J. Warren, T. A. Tronic, and J. M. Mayer, Chem. Rev. **110**, 6961 (2010).

[23] C. Bronner, and O. S. Wenger, J. Phys. Chem. Lett. **3**, 70 (2012).

[24] C. Bronner, and O. S. Wenger, Inorg. Chem. **51**, 8275 (2012).

[25] N. Barman, D. Singha, and K. Sahu, J. Phys. Chem. A **117**, 3945 (2013).

[26] T. Hossen, and K. Sahu, J. Phys. Chem. A **122**, 2394 (2018).

[27] J. Schirmer, Phys. Rev. A **26**, 2395 (1982).

[28] A. B. Trofimov, G. Stelter, and J. Schirmer, J. Chem. Phys. **111**, 9982 (1999).

[29] C. Hättig, Adv. Quant. Chem. **50**, 37 (2005).

[30] A. Dreuw, and M. Wormit, WIREs Comput. Mol. Sci. **5**, 82 (2014).

[31] R. S. Hosmane, M. A. Rossman, and N. J. Leonard, J. Am. Chem. Soc. **104**, 5497 (1982).

[32] M. Shahbaz, S. Urano, P. R. LeBreton, M. A. Rossman, R. S. Hosmane, and N. J. Leonard, J. Am. Chem. Soc. **106**, 2805 (1984).

[33] P. de Silva, J. Phys. Chem. Lett. **10**, 5674 (2019).

[34] H. Kollmar, and V. Staemmler, Theoret. Chim. Acta (Berl.) **48**, 223 (1978).

[35] J. Ehrmaier, E. J. Rabe, S. R. Pristash, K. L. Corp, C. W. Schlenker, A. L. Sobolewski, and W. Domcke, J. Phys. Chem. A **123**, 8099 (2019).

[36] W. Leupin, and J. Wirz, J. Am. Chem. Soc. **102**, 6068 (1980).

[37] W. Leupin, D. Magde, G. Persy, and J. Wirz, J. Am. Chem. Soc. **108**, 17 (1986).

[38] D. Gust, T. A. Moore, and A. L. Moore, Faraday Discuss. **155**, 9 (2012).

[39] B. Limburg, E. Bouwman, and S. Bonnet, Coord. Chem. Rev. **256**, 1451 (2012).

[40] Y. Cui, Z. Ding, P. Liu, M. Antonietti, X. Fu, and X. Wang, Phys. Chem. Chem. Phys. **14**, 1455 (2012).

[41] Y. Chen, J. Zhang, M. Zhang, and X. Wang, Chem. Sci. **4**, 3244 (2013).

[42] H. Wang, S. Jiang, S. Chen, D. Li, X. Zhang, W. Shao, X. Sun, J. Xie, Z. Zhao, and Q. Zhang, Adv. Mater. **28**, 6940 (2016).

[43] W. Domcke, D. R. Yarkony, and H. Köppel, *Conical Intersections: Electronic Structure, Dynamics and Spectroscopy* (World Scientific, Singapore, 2004).

[44] N. Makri, and W. H. Miller, J. Chem. Phys. **91**, 4026 (1989).

[45] T. Hammer, M. D. Coutinho-Neto, A. Viel, and U. Manthe, J. Chem. Phys. **131**, 224109 (2009).

[46] T. Hayashi, and S. Mukamel, J. Phys. Chem. A **107**, 9113 (2003).

[47] M. Savarese, P. A. Netti, C. Adamo, N. Rega, and I. Ciofini, J. Phys. Chem. B **117**, 16165 (2013).

[48] I. Polyak, C. S. M. Allan, and G. A. Worth, J. Chem. Phys. **143**, 084121 (2015).

[49] T. Raeker, and B. Hartke, J. Phys. Chem. A **121**, 5967 (2017).

[50] E. J. Rabe, K. L. Corp, X. Huang, J. Ehrmaier, R. G. Flores, S. L. Estes, A. L. Sobolewski, W. Domcke, and C. W. Schlenker, J. Phys. Chem. C **123**, 29580 (2019).

[51] M. G. D. Nix, A. L. Devine, B. Cronin, R. N. Dixon, and M. N. R. Ashfold, J. Chem. Phys. **125**, 133318 (2006).

[52] J. Ehrmaier, X. Huang, E. J. Rabe, K. L. Corp, C. W. Schlenker, A. L. Sobolewski, and W. Domcke, J. Phys. Chem. A **124**, 3698 (2020).

[53] X. Pang, C. Jiang, W. Xie, and W. Domcke, Phys. Chem. Chem. Phys. **21**, 14073 (2019).

[54] X. Huang, J.-P. Aranguren, J. Ehrmaier, J. A. Noble, W. Xie, A. L. Sobolewski, C. Dedonder-Lardeux, C. Jouvet, and W. Domcke, Phys. Chem Chem. Phys. **22**, 12502 (2020).

[55] N. Ullah, S. Chen, Y. Zhao, and R. Zhang, J. Phys. Chem. Lett. **10**, 4310 (2019).

[56] K. L. Corp, E. J. Rabe, X. Huang, J. Ehrmaier, M. E. Kaiser, A. L. Sobolewski, W. Domcke, and C. W. Schlenker, J. Phys. Chem. C **124**, 9151 (2020).

[57] J. Cabanillas-Gonzalez, G. Grancini, and G. Lanzani, Adv. Mater. **23**, 5468 (2011).

[58] S. D. Dimitrov, A. A. Bakulin, C. B. Nielsen, B. C. Schroeder, J. Du, H. Bronstein, I. McCulloch, R. H. Friend, and J. R. Durrant, J. Am. Chem. Soc. **134**, 18189 (2012).

[59] A. A. Bakulin, C. Silva, and E. Vella, J. Phys. Chem. Lett. **7**, 250 (2016).

[60] G. M. Paterno, L. Moretti, A. J. Barker, Q. Chen, K. Müllen, A. Narita, G. Cerullo, F. Scotognella, and G. Lanzani, Adv. Funct. Mater. **29**, 1805249 (2019).

[61] W. Domcke, J. Ehrmaier, and A. L. Sobolewski, ChemPhotoChem **3**, 10 (2019).

[62] G. Herzberg, *Molecular Spectra and Molecular Structure. Vol. I. Spectra of Diatomic Molecules* (Van Nostrand, New York, 1966),

[63] M. N. R. Ashfold, B. Cronin, A. L. Devine, R. N. Dixon, and M. G. D. Nix, Science **312**, 1637 (2006).

[64] M. N. R. Ashfold, G. A. King, D. Murdock, M. G. D. Nix, T. A. A. Oliver, and A. G. Sage, Phys. Chem. Chem. Phys. **12**, 1218 (2010).

[65] R. Montero, A. Peralta Conde, V. Ovejas, R. Martinez, F. Castano, and A. Longarte, J. Chem. Phys. **135**, 054308 (2011).

[66] G. M. Roberts, C. A. Williams, J. D. Young, S. Ullrich, M. J. Paterson, and V. G. Stavros, J. Am. Chem. Soc. **134**, 12578 (2012).

[67] G. M. Roberts, and V. G. Stavros, Chem. Sci. **5**, 1698 (2014).

[68] A. L. Sobolewski, W. Domcke, C. Dedonder-Lardeux, and C. Jouvet, Phys. Chem. Chem. Phys. **4**, 1093 (2002).

[69] J. H. Starcke, M. Wormit, and A. Dreuw, J. Chem. Phys. **130**, 024104 (2009).

[70] A. L. Sobolewski, and W. Domcke, Phys. Chem. Chem. Phys. **4**, 4 (2002).

[71] V. Balzani, L. Moggi, M. F. Manfrin, F. Boletta, and M. Gleria, Science **189**, 852 (1975).

[72] J. R. Bolton, Science **202**, 705 (1978).

[73] J. R. Bolton, S. J. Strickler, and J. S. Conolly, Nature **316**, 495 (1985).

[74] H. B. Gray, and A. W. Maverick, Science **214**, 1012 (1978).

[75] M. D. Archer, and J. R. Bolton, J. Phys. Chem. **94**, 8028 (1990).

[76] M. G. Walter, E. L. Warren, J. R. McKone, S. W. Boettcher, Q. Mi, E. A. Santori, and N. S. Lewis, Chem. Rev. **110**, 6446 (2010).

[77] J. R. Swierk, and T. E. Mallouk, Chem. Soc. Rev. **42**, 2357 (2013).

[78] M. D. Kärkäs, O. Verho, E. V. Johnston, and B. Åkermark, Chem. Rev. **114**, 11863 (2014).

[79] J. Schneider, M. Matsuoka, M. Takeuchi, J. Zhang, Y. Horiuchi, M. Anpo, and D. W. Bahnemann, Chem. Rev. **114**, 9919 (2014).

[80] J. D. Blakemore, R. H. Crabtree, and G. W. Brudvig, Chem. Rev. **115**, 12974 (2015).

[81] X. Luo, P. R. Fleming, and T. R. Rizzo, J. Chem. Phys. **96**, 5659 (1992).

[82] B. Rusic, J. Phys. Chem. A **119**, 7810 (2015).

[83] M. Teranishi, R. Hoshino, S. Naya, and H. Tada, Angew. Chem. Int. Ed. **55**, 12773 (2016).

[84] L. Yang, G. Dong, D. L. Jacobs, Y. Wang, L. Zang, and C. Wang, J. Catal. **352**, 274 (2017).

[85] G. Moon, M. Fujitsuka, S. K. Kim, T. Majima, X. Wang, and W. Choi, ACS Catal. **7**, 2886 (2017).

[86] A. Torres-Pinto, M. J. Sampaio, C. G. Silva, J. L. Faria, and A. M. T. Silva, Catalysts **9**, 990 (2019).

[87] H. Miyasaka, A. Tabata, S. Ojima, N. Ikeda, and N. Mataga, J. Phys. Chem. **97**, 8222 (1993).

[88] N. Mataga, Adv. Chem. Phys. **107**, 431 (1999).

[89] J. Waluk, Acc. Chem. Res. **36**, 832 (2003).

[90] A. L. Sobolewski, and W. Domcke, J. Phys. Chem. A **111**, 11725 (2007).

[91] D. R. Weinberg, C. J. Gagliardi, J. F. Hull, C. F. Murphy, C. A. Kent, B. C. Westlake, A. Paul, D. H. Ess, D. G. McCafferty, and T. J. Meyer, Chem. Rev. **112**, 4016 (2012).

[92] M. G. Kuzmin, I. V. Soboleva, V. L. Ivanov, E.-A. Gould, D. Huppert, and K. M. Solntsev, J. Phys. Chem. B **119**, 2444 (2015).

[93] S. M. Aspera, M. David, and H. Kasai, Japan. J. Appl. Phys. **49**, 115703 (2010).

[94] X. L. Wang, W. Q. Fang, H. F. Wang, H. Zhang, H. Zhao, Y. Yao, and H. G. Yang, J. Mater. Chem. A **1**, 14089 (2013).

[95] H.-Z. Wu, L.-M. Liu, and S.-J. Zhao, Phys. Chem. Chem. Phys. **16**, 3299 (2014).

[96] J. Wirth, R. Neumann, M. Antonietti, and P. Saalfrank, Phys. Chem. Chem. Phys. **16**, 15917 (2014).

[97] H. Ma, J. Feng, M. Wei, C. Liu, and Y. Ma, Nanoscale **10**, 15624 (2018).

[98] J. Sun, X. Li, and J. Yang, Nanoscale **10**, 3738 (2018).

[99] C.-Q. Xu, Y.-H. Xiao, Y.-X. Yu, and W.-D. Zhang, J. Mater. Sci. **53**, 409 (2018).

[100] P. Du, and R. Eisenberg, Energy Environ. Sci. **5**, 6012 (2012).

[101] S. Nagarajan, N. C. Skillen, F. Fina, G. Zhang, C. Random, N. A. Lawton, J. T. S. Irvine, and P. K. J. Robertson, J. Photochem. Photobiol. A **334**, 13 (2017).

[102] Y. Yang, J. Chen, Z. Mao, N. An, D. Wang, and B. D. Fahlmann, RSC Adv. **7**, 2333 (2017).

[103] J. Thiebaud, F. Thevenet, and C. Fittschen, J. Phys. Chem. C **114**, 3082 (2010).

[104] S. Tan, H. Feng, Y. Ji, Y. Wang, J. Zhao, B. Wang, Y. Luo, J. Yang, and J. G. Hou, J. Am. Chem. Soc. **134**, 9978 (2012).

[105] J. Zhang, and Y. Nosaka, J. Phys. Chem. C **117**, 1383 (2012).

[106] W. Kim, T. Tachikawa, G.-h. Moon, T. Majima, and W. Choi, Angew. Chem. Int. Ed. **53**, 14036 (2014).

[107] Y. Nosaka, and A. Y. Nosaka, Chem. Rev. **117**, 11302 (2017).